\documentclass[11pt]{article}
\usepackage{sint,macros,cite}
\usepackage{amssymb,amsmath}
\usepackage{epsfig}
\begin{document}
\begin{titlepage}

\renewcommand{\thefootnote}{\fnsymbol{footnote}}

\begin{flushleft}
\vbox{
\epsfxsize=4.0 true cm
\epsfbox{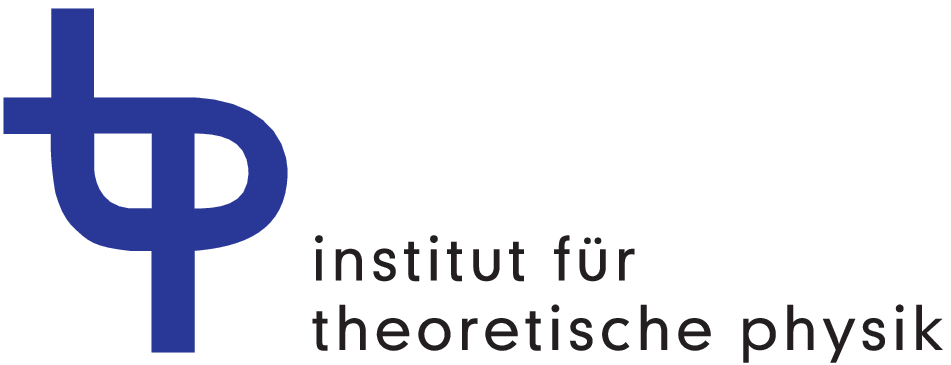}}
\end{flushleft}

\begin{flushright}
\vspace{-1.0cm}
MS-TP-03-14
\end{flushright}

\vskip 1.25cm
\begin{center}
{\Large\bf
Effective heavy-light meson energies\\[0.5ex] 
in small-volume quenched QCD
}
\end{center}
\vskip 0.875cm
\vbox{
\centerline{
\epsfxsize=2.5 true cm
\epsfbox{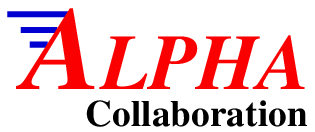}}
}
\vskip 1.0cm
\begin{center}
{\large
Jochen Heitger and 
Jan Wennekers\footnote[2]{Present address: 
                          DESY, Theory Group, Notkestrasse~85,
                          D-22603 Hamburg, Germany}
}
\vskip 1.0cm
Westf\"alische Wilhelms-Universit\"at M\"unster,
Institut f\"ur Theoretische Physik,\\
Wilhelm-Klemm-Strasse~9, D-48149 M\"unster, Germany
\vskip 1.5cm
{\bf Abstract}
\vskip 0.7ex
\end{center}

We study effective energies of heavy-light meson correlation functions 
in lattice QCD and a small volume of $(0.2\,\fm)^4$ to non-perturbatively 
calculate their dependence on the heavy quark mass in the continuum limit.
Our quenched results obtained here constitute an essential intermediate 
step of a first fully non-perturbative computation of the b-quark's mass 
in the static approximation that has recently been presented as an 
application of a new proposal to non-perturbatively renormalize the Heavy 
Quark Effective Theory.
The renormalization constant and the improvement coefficients relating the 
renormalized current and subtracted quark mass are determined in the 
relevant parameter region at weak couplings, which allows to perform the 
numerical simulations at several, precisely fixed values of the 
renormalization group invariant heavy quark mass in a range from 
$3\,\GeV$ to $15\,\GeV$.



\vskip 0.75cm
\vfill

\begin{center}
% \today
December 2003
\end{center}

\eject
\vfill
\eject

\renewcommand{\thefootnote}{\arabic{footnote}}

\end{titlepage}

\section{Introduction}
\label{Sec_intro}
During the last years, the elementary particle physics community has seen a
growing interest and activity in the study of B-meson physics and its 
phenomenology.
While on the experimental side this interest reflects in the various 
facilities that are currently running to explore CP-violation in the 
B-system \cite{Back:2003ty,Yamauchi:2003rw,Zoccoli:2003ih}, it is nourished
on the theoretical side by the demand to determine transition matrix elements 
of the effective weak Hamiltonian in order to interpret the experimental 
observations within (or beyond) the standard model and to (over-)constrain 
the unitarity triangle.
For the computations of such matrix elements between low-energy hadron states
to become valuable contributions in this field, they have to be carried out
non-perturbatively, which is the domain of QCD on the lattice.
However, in contrast to light quarks which as widely spread objects are
predominantly exposed to large-volume limitations, heavy quarks are extremely 
localized ($1/\mb\simeq 1/(4\,\GeV)\simeq 0.04\,\Fm$) and thus also require 
very fine lattice resolutions, because otherwise one would face huge 
discretization errors. 
Mainly for this reason, realistic simulations of heavy-light systems 
involving a b-quark (even in the quenched approximation) are impossible so 
far \cite{lat02:bphys}.

A theoretically very appealing way out of this restriction is to recourse to 
the \emph{Heavy Quark Effective Theory 
(HQET)} \cite{stat:eichten,stat:eichhill1}.
This comes at a prize, though.
As a consequence of its different renormalization properties, physical 
quantities deriving from expectation values calculated in the effective 
theory are affected by power-law divergences in the lattice spacing that can
not be subtracted perturbatively in a clean way:
the continuum limit ceases to exist, unless the theory is renormalized 
non-perturbatively \cite{Maiani:1992az}.

Only recently, an approach to overcome these deficiencies has been developed
from a solution for the problem of a completely non-perturbative computation 
of the b-quark mass in the static approximation \emph{including} the 
power-divergent subtraction \cite{lat01:mbstat} to a new method addressing 
the general class of renormalization problems in HQET \cite{mbstat:pap1}.
At its root stands the insight that these power-law divergences can be 
removed by a \emph{non-perturbative matching procedure to relativistic QCD in 
a finite volume}.
In fact, the \emph{smallness} of the physical volume represents a 
characteristic feature in this strategy --- since only then one is capable to 
incorporate and simulate the b-quark as a relativistic fermion --- and
motivates the investigation of QCD observables in a small-volume setup
presented in this work.
Here we will concentrate on the non-perturbative heavy quark mass dependence 
of effective heavy-light meson energies in the continuum limit, whose 
numerical knowledge is crucial to apply the proposal of \Ref{mbstat:pap1} to 
the determination of the b-quark's mass in leading order of HQET.\footnote{
For another method to determine $\mb$ and $\Fb$, which also starts from 
lattice QCD in small volume but employs extrapolations of finite-volume 
effects in the heavy quark mass, see \Refs{mb:roma2,fb:roma2c}.
} 
An extension to also examine the mass dependence of a few more quantities,
which aims at quantitative non-perturbative tests of HQET by comparing static
results with those obtained along the large quark mass limit in small-volume 
QCD, is in progress \cite{QCDvsHQET:pap2}.

In \Sect{Sec_strat} we first recall the main ideas of the general matching
strategy between HQET and QCD of \Ref{mbstat:pap1} (\Sect{Sec_strat_svol}), 
then introduce our observables considered (\Sect{Sec_strat_latobs}) and
finally describe how these can be calculated as functions of the 
(renormalized) mass of the heavy quark by numerical 
simulations (\Sect{Sec_strat_mqdep}).
\Sect{Sec_res_coeffs} is devoted to some intermediate results on a 
renormalization constant and improvement coefficients that are needed to 
renormalize the heavy quark mass in the relevant parameter range.
Our central results on the mass dependence of the heavy-light meson energies
are discussed in \Sect{Sec_res_gamrel}, and we conclude in \Sect{Sec_concl}.

In the calculations reported here we still stay in the quenched 
approximation.
We want to emphasize, however, that this study of the heavy quark mass 
dependence of suitable observables (as an important part of the general 
non-perturbative approach to HQET of \cite{mbstat:pap1}) can be expected to
be numerically implementable as well for QCD with dynamical fermions at a 
tolerable computing expense, because on basis of the experiences made
in \Refs{alpha:Nf2,Nf2SF:algo} the use of the QCD Schr\"odinger functional
brings us in a favourable position where in physically small to intermediate 
volumes dynamical simulations are definitely easier than with the standard 
formulation.

\section{Computational strategy}
\label{Sec_strat}
For the rest of this paper we will focus on meson observables derived from
heavy-light correlation functions in finite volume and their dependence
on the heavy quark mass in the continuum limit.
Before we explain in detail how this dependence can be singled out in an
actual numerical computation, we want to clarify the special r\^{o}le of 
QCD in a \emph{finite} volume as a material component of the more general
idea advocated in \Ref{mbstat:pap1} to non-perturbatively renormalize HQET;
therefore, it suggests itself to briefly summarize this idea in the first 
subsection.
\subsection{From QCD in small volume 
            to non-perturbatively renormalized HQET}
\label{Sec_strat_svol}
A long-standing problem with lattice computations in HQET is the occurrence 
of power-law divergences during the renormalization process 
(cf.~\Refs{mbstat:dm_MaSa,mbstat:dm_DirScor,mbstat:dm_Trottier}), implied by 
the allowed mixings of operators of different dimensions coming with 
coefficients $\{c_k\}$ that contain inverse powers of the lattice 
spacing $a$.
At each order of the HQET expansion parameter ($1/m$, where $m$ is the
heavy quark mass), new such free parameters $c_k$ arise, which in principle 
are adjustable by a matching to QCD; but owing to incomplete cancellations 
when performing this matching only in perturbation theory, one is always 
left with a perturbative remainder that still stays divergent as 
$a\rightarrow 0$.
Therefore, the continuum limit does not exist.

Already HQET in leading order, the static approximation, exhibits this 
unwanted phenomenon.
In this case the kinetic and the mass terms in the static action mix under 
renormalization and give rise to a local mass counterterm 
$\delta m\propto 1/a$, the self-energy of the static quark, which causes
a linearly divergent truncation error if one relies on an only perturbative 
subtraction of this divergence.
A prominent example for a quantity suffering from it is the b-quark mass 
itself: past computations in the static approximation \cite{lat01:ryan} 
were limited to finite lattice spacings, and the continuum limit was 
impossible to reach.

A viable strategy to solve this severe problem of power divergences is 
provided by a \emph{non-per\-tur\-ba\-tive matching of HQET and QCD in 
finite volume} \cite{lat01:mbstat,mbstat:pap1}.
To integrate the present work into the broader context of \Ref{mbstat:pap1}, 
we reproduce the central line of reasoning here.
Let us consider QCD consisting of (generically $\nf-1$) light quarks and 
a heavy quark, typically the b-quark.
In the effective theory, the dynamics of the heavy quark is governed by 
the HQET action, which formally is an expansion in inverse powers of the 
heavy quark mass.
(For further details see e.g.~Section~2 of \Ref{mbstat:pap1}.)
On the renormalized level, the effective theory discretized on a lattice can
be defined in terms of parameters $\{c_k\}$ that comprise those specifying 
the light quark sector (e.g.~the bare gauge coupling, $g_0^2\equiv c_1$, and 
the masses of the light quarks) and coefficients of local composite fields
in the $1/m$--expansion of the lattice action, supplemented by further
coefficients belonging to local composite operators which will be needed 
when including their correlation functions into the renormalization program.  
In other words, if these parameters are chosen correctly, HQET and QCD are
expected to be equivalent in the sense that
$\PhiHQET(M)=\PhiQCD(M)+\Or(1/M^{n+1})$ holds for suitable observables 
$\Phi$ in both theories, where for simplicity only the dependence on the 
heavy quark mass, here represented by the scheme and scale independent 
(and thus theoretically unambiguous) renormalization group invariant quark 
mass, $M$, is stressed.
Now suppose that in some way the parameters of QCD have already been fixed 
to proper values.
Then the parameters $\{c_k\}$ in the effective theory may just be determined 
through its relation to QCD by requiring a set of matching conditions:
\be
\PhiHQET_k(L,M)=\PhiQCD_k(L,M)\,, \quad k=1,\ldots,N_n\,.
\label{cond_match}
\ee
In this equation, the index $k$ labels the elements of the parameter set 
$\{c_k\}$ defining the effective theory up to $1/M^{n+1}$--corrections
(where, for instance, the additive mass renormalization $\delta m$ mentioned 
above is amongst them), and the conditions (\ref{cond_match}) determine the 
$c_k$ for any value of the lattice spacing.
Moreover, we have already indicated the dependence of the observables 
$\Phi_k$ in \eq{cond_match} on another variable which will become important 
in the following: the linear extent $L$ of a \emph{finite} volume.  

To substantiate this $L$--dependence, we note that in order to circumvent 
the difficulties with the power-law divergences from the start, the 
matching equations (\ref{cond_match}) are understood as 
\emph{non-perturbative conditions} in which both sides are to be calculated
with the aid of numerical simulations. 
From the practical point of view, this in turn also means that one must be 
able to simulate the b-quark as a relativistic fermion.
Hence, the linear extent $L$ of the matching volume (i.e.~where 
\eq{cond_match} should hold) has to be chosen with care.
On the one hand, it should fulfill $LM\gg 1$ to apply HQET quantitatively 
on the l.h.s.~of (\ref{cond_match}), i.e.~to suppress $1/m$--corrections, 
and on the other hand one has to ensure $aM\ll 1$ on the r.h.s.~to treat the 
heavy quark flavour in the relativistic theory and avoid large lattice 
artifacts so that the continuum limit is under control.
Taking these constraints together, while at the same time keeping the number 
of lattice points manageable for present-day computers, one then ends up 
with a volume for imposing the matching conditions (\ref{cond_match}) that 
is physically small; in our application later it will turn out to be of the 
order of $(0.2\,\fm)^4$. 

Having highlighted QCD in finite volume as key ingredient for the practical 
realization of the non-perturbative matching, we close this subsection with 
a few remarks on the subsequent (but not less important) steps that are 
involved to adopt it as a general approach for non-perturbative computations 
using the lattice regularized HQET.
These steps, together with two applications of the proposal as a whole, are
also worked out in \Ref{mbstat:pap1}, which the reader should consult for a
thorough discussion.
\begin{itemize}
\item Rather than directly identifying the quantities $\Phi_k$ entering 
      (\ref{cond_match}) with physical, experimentally accessible 
      observables (by which one would sacrifice the predictability of the 
      effective theory), they must be properly chosen as renormalized
      quantities computable in the continuum limit of lattice QCD, which
      in turn necessitates the use of a small volume as explained before.
      Of course, their explicit form still depends on the application in 
      question.
      One may think e.g.~of hadronic matrix elements or, as in the
      computations reported in the following sections, effective masses
      (respectively, energies) that are deduced from the decay of two-point 
      heavy-light correlation functions.
\item Apparently, although these matching conditions to fix the parameters 
      of HQET are perfectly legitimate (since the underlying Lagrangian does 
      not `know' anything about the finite volume), one still has to make
      contact to a physical situation, where the interesting quantities of 
      the infinite-volume theory can be extracted at the end.
      Employing the same lattice resolution as in the small volume 
      computation, however, would again demand too many lattice points.
      This gap between the small volume with its fine lattice resolution on
      the one side and larger lattice spacings, and thereby also larger 
      physical volumes, on the other is bridged by a recursive finite-size 
      scaling procedure inspired by \Ref{alpha:sigma}.
      As a result, the $\PhiHQET_k$ are obtained at some larger volume of 
      extent $L=\Or(1\,\Fm)$, where the resolutions $a/L$ are such that at 
      the same lattice spacings (i.e.~at the same bare parameters) volumes 
      with $L\simeq 2\,\Fm$ --- to accommodate physical observables in the 
      infinite-volume theory --- are affordable.
\item Finally, the approach requires a physical, dimensionful input.
      This usually amounts to relate the observables $\PhiHQET_k$ of the 
      effective theory calculated in large volume to some experimental 
      quantity.
      Which quantity this actually might be has to be decided when a
      concrete application is addressed.
      (E.g., in the application to compute $\mbbar$, it is the B-meson 
      mass \cite{mbstat:pap1}.)
\end{itemize}
\subsection{Lattice setup and observables}
\label{Sec_strat_latobs}
In our investigation of QCD in a small volume we distinguish between a
light (`l') and a heavy (`h') quark flavour.
The lattice regularized theory is formulated in a Schr\"odinger functional
(SF) cylinder of extent $L^3\times T$ \cite{SF:LNWW,SF:stefan1}: the gluon 
and quark fields are subject to periodic (Dirichlet) boundary conditions in 
spatial (temporal) directions, and 
--- if not explicitly stated otherwise --- we assume $T=L$ from now on.
In principle, the aforementioned matching strategy between HQET and QCD
is not restricted to the SF as the only possible finite-volume scheme to 
treat the involved heavy-light systems in the relativistic theory, but 
in this way its practical implementation will much profit from known
non-perturbative results on the renormalized quantities that will be needed
in the following.

\begin{figure}[htb]
\centering
\epsfig{file=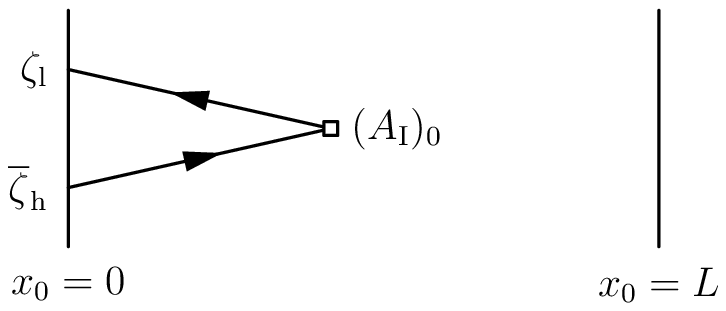,width=7.0cm}
\caption{{\fns
Illustration of the correlation function $\fa$, while in $\kv$ the 
insertion of $(\aimpr)_0$ is replaced by $(\vimpr)_k$.
}}\label{fig:fa}
\end{figure}

Starting from the $\Or(a)$ improved heavy-light axial and vector currents,
\bea
(\aimpr)_{\mu}(x) 
& = &
\lightb(x)\gamma_{\mu}\gamma_5\psi_{\rm h}(x)
+a\ca\half(\drv{\mu}+\drvstar{\mu})\left\{
\lightb(x)\gamma_5\psi_{\rm h}(x)\right\}\,,
\label{aimpr}\\
(\vimpr)_{\mu}(x) 
& = &
\lightb(x)\gamma_{\mu}\psi_{\rm h}(x)
+a\cv\half(\drv{\nu}+\drvstar{\nu})\left\{
i\,\lightb(x)\sigma_{\mu\nu}\psi_{\rm h}(x)\right\}\,,
\label{vimpr}
\eea
(where numerical values for the coefficients $\ca$ and $\cv$ in the 
quenched approximation are taken from \Refs{impr:pap3,lat97:marco}), 
we construct their correlation functions in the 
SF \cite{impr:pap1,impr:scalt} as
\bea
\fa(x_0) 
& = & 
-{a^6\over 2}\sum_{\vecy,\vecz}\mvl{
(\aimpr)_0(x)\,\zetabar_{\rm h}(\vecy)\gamma_5\zeta_{\rm l}(\vecz)}\,,
\label{fa}\\
\kv(x_0) 
& = & 
-{a^6 \over 6}\sum_{\vecy,\vecz,k}\mvl{
(\vimpr)_k(x)\,\zetabar_{\rm h}(\vecy)\gamma_k\zeta_{\rm l}(\vecz)}\,,
\label{kv}
\eea
schematically drawn in \Fig{fig:fa}.
%
%%%%%% figure: correlation function f_A
%
Based on these, we define $L$--dependent energies in both channels by 
combining the usual forward and backward difference operators to take the 
logarithmic derivatives
\bea
\gamps(L,M)
& \equiv &
-\frac{1}{2}\left(\partial_0+\partial_0^{\ast}\right)
\ln\left[\,\fa(x_0,M)\,\right]\,\Big|_{\,x_0=L/2}\,,
\label{gamps}\\
\gamv(L,M)
& \equiv &
-\frac{1}{2}\left(\partial_0+\partial_0^{\ast}\right)
\ln\left[\,\kv(x_0,M)\,\right]\,\Big|_{\,x_0=L/2}\,, \quad T=L\,,
\label{gamv}
\eea
in which all multiplicative renormalization factors drop out.
These observables also depend on the bare coupling, $g_0$, and --- as the
light quark mass is put to zero in the actual computations --- on the heavy
quark mass variable which, as already in \Sect{Sec_strat_svol}, one may
conveniently set to be the \emph{renormalization group invariant (RGI)} mass 
of the heavy quark, $M$.
Having in mind the corresponding correlators in the effective theory, where
in the static (i.e.~infinite-mass) limit the two currents fall together
owing to the heavy quark spin-symmetry, we also form their spin-averaged 
combination
\be
\gamav\equiv\frac{1}{4}\left(\gamps+3\gamv\right)\,,
\label{gamav}
\ee
which is expected to deviate from the static limit by the smallest 
$1/m$--corrections \cite{mbstat:pap1}. 
\subsection{Determination of the heavy quark mass dependence}
\label{Sec_strat_mqdep}
As mentioned before and detailed in \Refs{lat01:mbstat,mbstat:pap1}, the 
non-perturbative matching between HQET and QCD can be applied to determine 
the mass of the b-quark in the static approximation.
To achieve this goal, the quantities $\Phi_k$ in \eq{cond_match} have to be
identified with suitable observables that can be calculated by numerical
simulations.
Since we want to evaluate them in the continuum limit, we have to ensure a 
situation with fixed renormalized parameters, along which the limit 
$a/L\rightarrow0$ may be performed.
In regard of the (suppressed as well as indicated) dependences in the 
matching conditions (\ref{cond_match}), this means that for each value of 
$L/a$ we have to make particular choices for the gauge coupling and the 
quark masses in the light and heavy sectors, respectively. 
The first two of them are immediately offered by
\bea
\PhiHQET_1
& = &
\gbsq(L)\,\,\,=\,\,\,\mbox{constant}\,,
\label{cond_gbar}\\
\PhiHQET_2
& = &
\ml\,\,\,=\,\,\,0\,,
\label{cond_ml}
\eea
where $\gbsq$ denotes the renormalized finite-volume coupling in the SF
scheme that runs with the box size $L$ \cite{alpha:SU3} and 
$\ml\equiv\ml^{\rm PCAC}$ is the PCAC current quark mass of the light
flavour.
These conditions allow to readily benefit from earlier work, because when
fixing $\gbsq$ to one of the values used for the (quenched) computation of 
the non-perturbative quark mass renormalization in \cite{msbar:pap1}, they
translate into triples $(L/a,\beta=6/g_0^2,\kapl)$ that can directly be 
taken over from this reference.\footnote{
In this case the lattice action's hopping parameter associated with the
light quark flavour, $\kapl$, just equals the critical hopping parameter,
$\kapc$, as a consequence its very definition: 
$\ml(\kapc)=0$, see \Ref{msbar:pap1}.
}

For the computation of the b-quark's mass in leading order of HQET one yet 
needs one more condition in order to fix the parameter $a\delta m$ in the 
static Lagrangian. 
Following \cite{lat01:mbstat,mbstat:pap1}, the most natural candidates for
this purpose are now the pseudoscalar and spin-averaged energies of the 
form (\ref{gamps}) and (\ref{gamav}) introduced in \Sect{Sec_strat_latobs}.
For $k=3$, \eq{cond_match} then turns into
\be
L\,(\gamstat + m)\,\equiv\,
\PhiHQET_3(L,M)\,=\,\PhiQCD_3(L,M)
\,\equiv\, L\,\gamx\,,
\label{cond_gam} 
\ee
where $\gamx=\gamx(L,M)$, ${\rm X}={\rm PS},{\rm av}$, are defined in 
\emph{QCD and small volume with a relativistic b-quark} and carry the 
entire quark mass dependence.
On the l.h.s., $\gamstat$ denotes the analogue of $\gamx$ in the effective 
theory, where the heavy quark entering the correlator (\ref{fa}) is treated
in the static approximation, and implicitly contains the problematic,
linearly divergent mass counterterm $a\delta m$.
The matching equation (\ref{cond_gam}) may be connected to a physical 
quantity in large volume (the B-meson mass \cite{mbstat:pap1}) by 
finite-size scaling (cf.~\Sect{Sec_strat_svol}).
But without going into further, unnecessary details we only note that this 
yields a (dimensionless) equation, which apart from $\gamx(L,M)$ 
only contains energy differences computable in the static theory 
--- such that the counterterm $a\delta m$ cancels out --- and the 
experimental mass of the B-meson.
Since all pieces in that equation can be evaluated non-perturbatively and in
the continuum limit, the quantitative knowledge of $\gamx(L,M)$ as a 
function of $M$ in the relevant quark mass region finally allows to solve it 
for the desired numerical value of the non-perturbatively renormalized RGI 
mass of the b-quark, $\Mb$, in the static approximation \cite{mbstat:pap1}.

So the task is here to non-perturbatively determine the heavy quark mass 
dependence of the effective heavy-light meson energies $\gamx(L,M)$.
It thus remains to discuss how we fix the linear size of the finite volume, 
$L=L_0$, and a set of dimensionless quark mass values,
\be
z\equiv L_0M=\mbox{constant}\,,
\label{cond_z}
\ee
at which the numerical computation of $\gamx$ for various lattice spacings 
takes place in practice.

We answer this question again on basis of the present numerical knowledge in 
the framework of the SF.
Among the available constant values of the renormalized SF coupling that
belong to different sets of triples $(L/a,\beta,\kapl)$ known
from \Ref{msbar:pap1} we fix, as was initially required by the condition 
(\ref{cond_gbar}),
\be
\gbsq(L_0/2)=1.8811\,,
\label{cond_gbarL0h}
\ee
where the (inverse) lattice spacing varies within $6 \leq L_0/(2a) \leq 16$.
This choice is particularly convenient, because from the same reference
one then finds that $L_0$ is given in units of a certain maximal size, 
$\lmax$, of the SF box, namely
\be
L_0=\frac{\lmax}{2}=0.36\,r_0\approx 0.18\,\Fm\,,
\label{choice_L0}
\ee
which is implicitly defined by $\gbsq(\lmax)=3.48$ and, through its relation 
to the hadronic radius $r_0=0.5\,\Fm$ \cite{pot:r0}, also known in physical 
units: $\lmax/r_0=0.718(16)$ \cite{pot:r0_ALPHA}.
Indeed, from a closer inspection of the parameters corresponding to the
condition (\ref{cond_gbarL0h}) (see \Tab{tab:param}) one may convince 
oneself that an accompanying volume of some small extent 
as (\ref{choice_L0}) is a sensible compromise between the associated 
resolutions being small enough to comfortably accommodate a propagating 
(albeit the system size confining) b-quark and being able to consider heavy 
quark masses large enough to suppress $\Or(1/z)$--corrections 
(i.e.~$z\gg1$ but not too large to avoid a substantial enhancement of cutoff 
effects) as well as to cover the physical RGI b-quark mass scale itself.

In order to satisfy \eq{cond_z} for the intended numerical values $z$ while 
$L/a$ and $\beta$ are changed to approach the continuum limit, we still have 
to relate the RGI mass of the heavy quark, $M$, to the corresponding hopping 
parameter $\kaph$ as the simulation parameter for the bare heavy quark mass
in the lattice action.
One possibility would be to employ the ($\Or(a)$ improved) relation between 
the RGI and the bare PCAC quark mass, $\mh$, viz.
\be
M=
\frac{M}{\mhbar(\mu_0)}\,
\frac{\za(g_0)(1+\ba a\mqh)}{\zp(g_0,L_0)(1+\bP a\mqh)}
\times\mh\,+\,\Or\left(a^2\right)\,,\quad \mu_0=1/L_0\,,
\label{M_mpcac}
\ee
where $\mhbar(\mu_0)$ denotes the running heavy quark mass renormalized at 
the scale $\mu_0=1/L_0$ in the SF scheme and (the inverse of) its 
\emph{ratio} to the RGI mass $M$, the first factor on the r.h.s., is 
inferable from the literature as will be elaborated on after \eq{def_h0}. 
The numerator and denominator in the second factor account for the 
renormalization of the axial current and the pseudoscalar density in the 
$\Or(a)$ improved theory, respectively.
Here, the subtracted bare heavy quark mass is defined as usual by 
$\mqh=\mzh-\mc$ with $\mc$ the critical value of the bare quark mass $\mzh$, 
which in terms of the hopping parameters is given by
$a\mqh=\half(\kaph^{-1}-\kapc^{-1})$.
Although the involved renormalization factors are in principle known 
from \Refs{msbar:pap1,impr:babp} in the quenched case, this would demand 
additional simulations to appropriately tune $\kaph$ 
--- and thereby the PCAC mass $\mh$ as a secondary quantity composed of 
correlation functions (cf.~\eq{mpcac_x0}) --- until the 
condition (\ref{cond_z}) is met with sufficient precision.

However, a safer (and also more economic) way to estimate $\kaph$ such that 
a constant $z=L_0M$ can be enforced in advance \emph{without} any additional 
tuning runs is as follows.
Recall that alternatively the renormalized quark mass can be written in 
terms of the $\Or(a)$ improved subtracted bare quark mass, $\mqhtil$, as
\be
\mhbar=\zm\,\mqhtil\,,\quad
\mqhtil=\mqh(1+\bm a\mqh)\,,
\label{def_mhbar}
\ee
where non-perturbative estimates on the improvement coefficient $\bm$ and
the finite combination of renormalization constants
\be
Z(g_0)\equiv\frac{\zm(g_0,L)\zp(g_0,L)}{\za(g_0)}
\label{def_Z}
\ee
have been published in \Ref{impr:babp} for the quenched approximation.
The latter identity may then be used to eliminate the (in the SF scheme
unknown) renormalization factor $\zm$ in \eq{def_mhbar} in favour of 
numerically known ones.
The decomposition of the RGI heavy quark mass analogous to (\ref{M_mpcac})
now assumes the form:
\be
M=
\frac{M}{\mhbar(\mu_0)}\,
\frac{Z(g_0)\za(g_0)}{\zp(g_0,L_0)}
\times\mqh(1+\bm a\mqh)\,+\,\Or\left(a^2\right)\,,\quad \mu_0=1/L_0\,.
\label{M_mqtil}
\ee
The last piece to be addressed in this equation is the universal, 
regularization independent ratio of the RGI quark mass to the renormalized
mass at fixed renormalization scale, which for later reference we call
\be
h(L_0)\equiv
\frac{M}{\mhbar(\mu_0)}\,,\quad \mu_0=1/L_0\,.
\label{def_h0}
\ee
In a mass independent renormalization scheme such as the SF it is flavour 
independent and, according to \Ref{msbar:pap1}, can be expressed as
\be
h(L_0)=
\frac{\zp(L_0)}{\zp(2^{-6}\lmax)}\,\times\,
\left[\,2b_0\gbsq(\mu)\,\right]^{-d_0/(2b_0)}
\exp\left\{-\int_{0}^{\gbar(\mu)}\rmd g
\left[\,\frac{\tau(g)}{\beta(g)}-\frac{d_0}{b_0g}\,\right]\right\}
\label{expr_h0}
\ee
with $L_0=\lmax/2$, $\mu=2^6/\lmax$ and $b_0,d_0$ the leading-order 
coefficients in the perturbative expansions of the renormalization group 
functions of the running coupling and quark mass, $\beta(\gbar)$ and 
$\tau(\gbar)$.
In the ratio of $\zp$--factors we encounter a scale evolution through 
changes by finite step sizes in the SF renormalization scale $L=1/\mu$
that has been non-perturbatively computed in \cite{msbar:pap1}, whereas the 
remainder lives in the high-energy regime where the coupling is reasonably
small to evaluate it in perturbation theory. 
Since in \eq{choice_L0} we just chose the linear extent $L_0$ of the small 
volume, where the non-perturbative matching to the relativistic theory is to 
be performed, as a proper multiple of the scale $\lmax$, a numerical 
estimate for $h(L_0)$ in the continuum limit, \eq{def_h0}, can be directly 
extracted from the results already published in \cite{msbar:pap1} and will 
be quoted in \Sect{Sec_res_gamrel} later on.

Our simulation parameters, which fix via \eqs{cond_gbar}, (\ref{cond_ml}) 
and (\ref{cond_gbarL0h}) the physics in the relativistic sector and via 
\eq{choice_L0} the size of the matching volume where to determine the 
non-perturbative heavy quark mass dependence of our observables, are 
summarized in \Tab{tab:param}.
%
%%%%%% table: parameter sets
%
\begin{table}[htb]
\centering
\vspace{0.25cm}
\begin{tabular}{ccccccccr@{.}l}
\hline \\[-2.0ex]
  set && $L/a$ & $\beta=6/g_0^2$ && $\kapl$ & $\gbsq(L/2)$ & $\gbsq(L)$ 
& \multicolumn{2}{c}{$\zp$} \\[1.0ex]
\hline            \\[-2.0ex]
  A && $12$ & $7.4082$ && $0.133961(8)$ & $1.8811(22)$ & $2.397(17)$ 
& $0$&$6764(6)$   \\
  B && $16$ & $7.6547$ && $0.133632(6)$ & $1.8811(28)$ & $2.393(18)$ 
& $0$&$6713(8)$   \\
  C && $20$ & $7.8439$ && $0.133373(2)$ & $1.8811(22)$ & $2.379(22)$ 
& $0$&$6679(8)$   \\
  D && $24$ & $7.9993$ && $0.133159(4)$ & $1.8811(38)$ & $2.411(20)$ 
& $0$&$6632(8)$   \\ 
  E && $32$ & $8.2415$ && $0.132847(3)$ & $1.8811(99)$ & $2.397(52)$ 
& $0$&$6575(13)$ \\[1.0ex]
\hline \\[-2.0ex]
\end{tabular}
\caption{{\fns
Our parameter sets that refer to the light quark sector and have fixed
SF coupling, $\gbsq(L/2)=1.8811$.
The parameters $L/a$, $\beta$ and $\kapl\equiv\kapc$ of A, B, D and E are
taken over from \protect\cite{msbar:pap1} without changes, whereas we 
performed new simulations to add with C a further lattice resolution for 
this work.
Moreover, the renormalization constants $\zp=\zp(g_0,L/a)$ differ from those
of \protect\cite{msbar:pap1} in that they have been computed in the context 
of \protect\cite{zastat:pap3} using the two-loop 
value \protect\cite{impr:ct_2loop} for the boundary improvement 
coefficient $\ct$.
}}\label{tab:param}
\end{table}
Values for the other quantities, which have to be inserted into \eq{M_mqtil}
in order to obey the condition (\ref{cond_z}) of constant $z=L_0M$, will be
specified when we come to describe the actual calculation of the 
$z$--dependence of $\gamx=\gamx(L_0,M)$, ${\rm X}={\rm PS},{\rm av}$,
in \Sect{Sec_res_gamrel}.

At this stage we still want to direct the reader's attention to an issue 
that can already be foreseen to become a potential limitation in any 
application of the numerically computed quark mass dependence:
even with \eq{M_mqtil}, the dimensionless mass parameter $z$ can only be 
fixed through quantities known to a certain precision.
In particular, the improvement coefficient $\bm$ and the renormalization 
factor $Z$ are only available from \Ref{impr:babp}\footnote{
The $\beta$--range considered in \cite{impr:babp} was chosen to cover
values commonly used to simulate $\Or(a)$ improved quenched QCD in 
physically large volumes.
Similarly, the results from determinations through chiral Ward identities 
with mass non-degenerate quarks \cite{impr:roma2_1,impr:losalamos} refer to 
this region and are not applicable here, too.
}
for $6.0\le\beta\le 6.756$, which are significantly lower than the $\beta$-s
of \Tab{tab:param} we are interested in.
It hence appears quite difficult to reliably guess from the numbers in 
\cite{impr:babp} values for $\bm$ and $Z$ in the $\beta$--region relevant 
here and, even more, to quantitatively assess the additional error 
contribution those estimates would then be afflicted with.  
Since this constitutes a dominant source of uncertainty, which would also
propagate into any quantity that explicitly derives from the $z$--dependence
of the observables studied here --- such as, for instance, the b-quark mass
through the matching of HQET and QCD \cite{mbstat:pap1} ---, we first 
determine $\bm$ and $Z$ exactly for the $\beta$--values of \Tab{tab:param}
and also improve their numerical precision.

\section{Results}
\label{Sec_res}
We now present the numerical results of our (quenched) computation of the
heavy-light meson energies 
$\gamx(L_0,M)$, ${\rm X}={\rm PS},{\rm av}$, which --- as outlined in the 
previous subsection --- basically consists of two parts: 
first, we determine $\bm$ and $Z$ in the $\beta$--range relevant for a
matching in physically small volume in order to be able to fix the RGI heavy 
quark mass to a set of desired values $z=L_0M$ with sufficient precision, 
and second, the $L_0\gamx(L_0,M)$ are calculated at these values of $z$ in 
the continuum limit.
This will then allow for smooth representations of the energies as functions 
of $z$, which eventually may be used to interpolate them to the b-quark 
scale, $\zb=L_0\Mb$ \cite{mbstat:pap1}. 
\subsection{Coefficients $\boldsymbol{\ba-\bP}$, $\boldsymbol{\bm}$ and 
            renormalization factor $\boldsymbol{Z}$ for 
            $\boldsymbol{7.4\lesssim\beta\lesssim 8.2}$}
\label{Sec_res_coeffs}
Here we proceed in the spirit of \Ref{impr:babp}, where the idea of 
\emph{imposing improvement conditions at constant physics} was advocated.
In that work, which considered a range $6.0\le\beta\le 6.756$ of bare 
couplings commonly used in large-volume simulations, this was realized by 
keeping constant the ratios $L/r_0$ and $T/L=3/2$ (supplemented by the 
SF-specific choices $C=C'=0$ and $\theta=0.5$). 
However, despite $\ln(a/r_0)$ as a function of $\beta$ needed to fix $L/r_0$ 
to some suitable value for given $L/a$ is available for 
$5.7\le\beta\le 6.92$ \cite{pot:r0_silvia} and even 
beyond \cite{pot:r0_largebeta}, the condition of constant $L/r_0$ can not 
be transfered to the present situation.
The reason lies in the fact that for $\beta$--values in the range we are 
interested in, $7.4\lesssim\beta\lesssim 8.2$, enforcing an improvement 
condition such as $L/r_0=\Or(1)$ would lead to prohibitively large values of 
$L/a$ in the simulations.
Therefore, to replace the latter, we exploit one of the already built in 
elements of the matching strategy between HQET and QCD as explained in the 
foregoing section: namely, as for the computation of the energies $\gamx$ on 
the QCD side we have to work along a line of constant physics in bare 
parameter space anyway, we can directly adopt the pairs $(L/a,\beta)$ 
of \Tab{tab:param}, which correspond to $L/\lmax=1/2=0.36r_0/\lmax$ and 
simultaneously to a constant renormalized coupling of $\gbsq(L/2)=1.8811$ 
(see \eqs{choice_L0} and (\ref{cond_gbarL0h})).
Within the present application, this constitutes a much more natural and 
equally admissible choice of improvement condition and thereby, in the same 
way as in \Ref{impr:babp}, the improvement coefficients $\ba-\bP$ and $\bm$ 
as well as the renormalization constant $Z$ become smooth functions of 
$g_0^2$ but exactly in the region where they are needed.

Following \cite{impr:babp} --- and also referring to this reference for any
unexplained details ---, the improvement coefficient $\bm$ and the 
renormalization constant $Z$ (as well as the difference of coefficients 
$\ba-\bP$, though it does not enter the subsequent computations
in \Sect{Sec_res_gamrel}) can be determined by studying QCD with 
non-degenerate quarks. 
This is particularly advantageous in case of the quenched approximation at 
hand, since then the structure of the $\Or(a)$ improved theory stays quite 
simple. 
For instance, the improvement of the off-diagonal bilinear fields 
$X^{\pm}=X^1\pm i\,X^2$, $X=A_{\mu},P$, emerging as a consequence of the 
broken isospin symmetry, is the same as in the degenerate case, except that
the $b$--coefficients now multiply cutoff effects proportional to the
average $\half(a\mqi+a\mqj)$ of the subtracted bare quark masses, 
$\mqi=\mzi-\mc$, which themselves are separately improved for each quark 
flavour:
\be
\mqitil=\mqi(1+\bm a\mqi)\,.
\ee 
(Here and below the indices $i,j$ label the different quark flavours.)
Identifying, for instance, the flavours in the isospin doublet with a light
and a heavy quark as before, the corresponding PCAC relation reads
\be
\Pmu A_{\mu}^{\pm}(x)=(\ml+\mh)P^{\pm}(x)\,,
\ee
and the renormalization constants $\za$ and $\zp$ that come into play
upon renormalization are just those known in the theory with two 
mass degenerate quarks.

Accordingly, the fermionic correlation functions defined in the SF and
involving the axial current and the pseudoscalar density \cite{impr:pap1}
generalize to $\fa^{ij}(x_0)=-\frac{1}{2}\mvl{A_0^+(x)O^-}$ and 
$\fp^{ij}(x_0)=-\frac{1}{2}\mvl{P^+(x)O^-}$, with pseudoscalar boundary
sources decomposed as $O^{\pm}=O^{1}\pm i\,O^{2}$ where
$O^{a}=a^6\sum_{\mby,\mbz}\zeb(\mby)\gfv\,\gen\,\zeta(\mbz)$.
Then the improved bare PCAC (current) quark masses\footnote{
This expression for the PCAC masses is only $\Or(a)$ improved up to a factor
$1+\half(\ba-\bP)(a\mqi+a\mqj)$ for quark mass dependent cutoff effects. 
}
as functions of the timeslice location $x_0$ are given by
\be
m_{ij}(x_0;L/a,T/L,\theta)=
\frac{\Sz\fa^{ij}(x_0)+a\ca\drvstar{0}\drv{0}\fp^{ij}(x_0)}
{2\,\fp^{ij}(x_0)}\,,
\label{mpcac_x0}
\ee
where only here we explicitly indicate their additional dependence on 
$L/a$, $T/L$ and the periodicity angle $\theta$ of the fermion fields.
In the degenerate case, $i=j$, the correlators assume the standard form
as introduced earlier \cite{impr:pap1,impr:scalt}, and $m_{ij}$ just
reduces to the current quark mass of a single quark flavour that is prepared 
by a corresponding choice of equal values for the associated hopping 
parameters, $\kappa_i=\kappa_j$.
Also the precise definition of the lattice derivatives in \eq{mpcac_x0}
matters.
As it is written there, $\Sz=\half(\drv{0}+\drvstar{0})$ denotes the average
of the usual forward and backward derivatives, but as in \Ref{impr:babp} we 
have employed the improved derivatives 
\be
\Sz\rightarrow
\Sz\left(1-{\T \frac{1}{6}}\,a^2\drvstar{0}\drv{0}\right)\,,\quad
\drvstar{0}\drv{0}\rightarrow
\drvstar{0}\drv{0}\left(1-{\T \frac{1}{12}}\,a^2\drvstar{0}\drv{0}\right)
\ee
as well, which (when acting on smooth functions) have $\Or(g_0^2a^2,a^4)$ 
errors only.

To enable their numerical calculation, the desired coefficients $\ba-\bP$, 
$\bm$ and the finite factor $Z=\zm\zp/\za$ have to be isolated.
This can be achieved by virtue of the identity
\bea
m_{ij}
& = &
Z\,\Big[\,
\half\left(\mqi+\mqj\right)+\half\,\bm\left(a\mqi^2+a\mqj^2\right)
\nonumber\\
&   &
\hspace{3.22cm}-\,\quart\left(\ba-\bP\right)a\left(\mqi+\mqj\right)^2
\,\Big]\,+\,\Or\left(a^2\right)\,,
\label{mpcac_mq}
\eea
which is obtained if one equates the two available expressions for the
$\Or(a)$ improved renormalized quark masses, i.e.~in terms of the current
quark and the subtracted bare quark masses (as they appear for the 
degenerate case in \eq{M_mpcac} and thereafter).
Forming ratios of suitable combinations of degenerate and non-degenerate 
current quark masses in the representation (\ref{mpcac_mq}) allows to
derive the following direct estimators for $\ba-\bP$, $\bm$ 
and $Z$ \cite{impr:babp}:
\bea
R_{\rm AP}
& = &
\frac{2\,(2m_{12}-m_{11}-m_{22})}
{(m_{11}-m_{22})(am_{{\rm q},1}-am_{{\rm q},2})}
\,\,\,=\,\,\,\ba-\bP\,+\,\Or\left(am_{{\rm q},1}+am_{{\rm q},2}\right)\,,
\label{estim_bap}\\
R_{\rm m}
& = &
\frac{4\,(m_{12}-m_{33})}
{(m_{11}-m_{22})(am_{{\rm q},1}-am_{{\rm q},2})}
\,\,\,=\,\,\,\bm\,+\,\Or\left(am_{{\rm q},1}+am_{{\rm q},2}\right)\,,
\label{estim_bm}
\eea
with $m_{0,3}=\half(m_{0,1}+m_{0,2})$, apart from other quark mass 
independent lattice artifacts of $\Or(a)$.
For the renormalization constant $Z$ an analogous expression holds even up
to $\Or(a^2)$ corrections,
\be
R_{Z}\,\,\,=\,\,\,
\frac{m_{11}-m_{22}}{m_{{\rm q},1}-m_{{\rm q},2}}
\,+\,(\ba-\bP-\bm)(am_{11}+am_{22})
\,\,\,=\,\,\,Z\,+\,\Or\left(a^2\right)\,,
\label{estim_Z}
\ee
provided that the correct value for $\ba-\bP-\bm=R_{\rm AP}-R_{\rm m}$ 
(only involving correlation functions with mass degenerate quarks) is 
inserted.

%
%%%%%% table: results on b_A-b_P, b_m and Z for 1st z-choice
%
\begin{table}[htb]
\centering
\begin{tabular}{cccccccc}
\hline \\[-2.0ex]
set && $\kaph$ & $L\mh$ && $\ba-\bP$ & $\bm$ & $Z$ \\[1.0ex]
\hline \\[-2.0ex]
  A && $0.132728$ & $0.4778(7)$ && $-0.0008(14)$ & $-0.6217(17)$ 
& $1.0941(3)$ \\
  B && $0.132711$ & $0.4621(7)$ && $-0.0059(22)$ & $-0.6218(27)$ 
& $1.0916(3)$ \\
  C && ---        & $0.4572(6)$ && $-0.0057(23)$ & $-0.6228(28)$ 
& $1.0900(3)$ \\
  D && $0.132553$ & $0.4539(5)$ && $-0.0072(21)$ & $-0.6260(27)$ 
& $1.0882(2)$ \\
  E && $0.132395$ & $0.4508(5)$ && $-0.0077(25)$ & $-0.6312(32)$ 
& $1.0859(2)$ \\[1.0ex]
\hline \\[-2.0ex]
\end{tabular}
\caption{{\fns
Numerical results on the improvement coefficients $\ba-\bP$ and $\bm$ and on
the renormalization constant $Z$, based on statistics varying between 
$\Or(900)$ measurements (A) and $\Or(200)$ measurements (E). 
These numbers refer to `choice~1' in \eq{choices_m}, where the heavy quark 
mass is kept at $L\mh\approx 0.5$, while $L\ml$ indeed turned out to be 
compatible with zero up to tiny deviations of $\Or(0.03)$ (A) and 
$\Or(0.01)$ (sets B -- E).
For C, we interpolated results obtained with $L/a=16,24$ to $L/a=20$.
Amongst others, $\bm$ and $Z$ are needed to fix the renormalized heavy quark 
mass in the simulations reported in \Sect{Sec_res_gamrel}.
}}\label{tab:bXzres_1}
\end{table}
Concerning the simulation parameters, we argued above that it is most 
natural to choose $L/a$ and $\beta$ exactly as listed in \Tab{tab:param} of 
the preceding section.
Moreover, values for the bare quark masses have to be selected.
Here we considered two pairs of values for them,
\bea
\mbox{choice~1}:
& \quad &
\mzl=\mc \,\Leftrightarrow\, L\ml\approx 0 \,,\quad L\mh\approx 0.5\,,
\nonumber\\
\mbox{choice~2}:
& \quad &
\mzl=\mc \,\Leftrightarrow\, L\ml\approx 0 \,,\quad L\mh\approx 2.6\,;
\label{choices_m}
\eea
to comply with the heavy-light notation before we identify the bare quark 
masses as $m_{0,1}=\mzl$ and $m_{0,2}=\mzh$ with associated PCAC masses 
$m_{11}\equiv\ml$ and $m_{22}\equiv\mh$.
With the first choice (corresponding to $z\approx 1$) $L\mh$ is close to the 
value used in \Ref{impr:babp}, which has the advantage that there this 
condition was also investigated in perturbation theory and the encountered
$\Or(a)$ ambiguities were not enormous.
On the contrary, the second one (corresponding to $z\approx 5$) is motivated 
by the typical quark mass region that we have to deal with when probing our 
non-degenerate, heavy-light quark system in the next subsection.
Maintaining these conditions to sufficient accuracy requires some prior 
simulations to tune the hopping parameter $\kaph$ belonging to the heavy 
flavour's bare mass appropriately, whereas by $\kapl=\kapc$ taken over from 
the known values of \Ref{msbar:pap1} the light quark mass is set to zero.
The final simulation parameters to determine the quantities 
$R_{\rm X}$, ${\rm X}={\rm AP},{\rm m},Z$, in question can be drawn from the
triples $(L/a,\beta,\kapl)$ of \Tab{tab:param} specifying the light quark 
sector (except that, only here, $T=3L/2$), while the bare heavy quark mass 
is set through $\kaph$ given in the second columns of \Tabs{tab:bXzres_1} 
and \ref{tab:bXzres_2}.
There one can also see that the bare current quark mass $L\mh$ has been 
fixed within a few percent to the values dictated by (\ref{choices_m}) to 
keep physics constant.\footnote{
Similar to the situation in \Ref{impr:babp}, this is to a good precision
equivalent to keeping fixed the corresponding renormalized masses
$L\mr=L\za m/\zp$, because also over our range of considered couplings 
the entering renormalization constant barely varies.
}

%
%%%%%% table: results on b_A-b_P, b_m and Z for 2nd z-choice
%
\begin{table}[htb]
\centering
\begin{tabular}{cccccccc}
\hline \\[-2.0ex]
set && $\kaph$ & $L\mh$ && $\ba-\bP$ & $\bm$ & $Z$ \\[1.0ex]
\hline \\[-2.0ex]
  A && $0.126040$ & $2.7100(6)$ && $0.0489(3)$ & $-0.5401(4)$ 
& $1.0855(2)$ \\
  B && $0.128028$ & $2.6112(6)$ && $0.0239(5)$ & $-0.5621(7)$ 
& $1.0867(2)$ \\
  C && ---        & $2.6709(5)$ && $0.0151(6)$ & $-0.5744(9)$ 
& $1.0867(2)$ \\
  D && $0.129595$ & $2.5456(5)$ && $0.0103(5)$ & $-0.5811(8)$ 
& $1.0859(1)$ \\
  E && $0.130246$ & $2.5035(5)$ && $0.0051(6)$ & $-0.5927(9)$ 
& $1.0845(1)$ \\[1.0ex]
\hline \\[-2.0ex]
\end{tabular}
\caption{{\fns
The same as in \Tab{tab:bXzres_1} but for `choice~2' of the heavy quark
mass: $L\mh\approx 2.6$.
}}\label{tab:bXzres_2}
\end{table}
%
%%%%%% figure: (b_A-b_P)(g_0)
%
\begin{figure}[htb]
\centering
\vspace{-1.625cm}
\epsfig{file=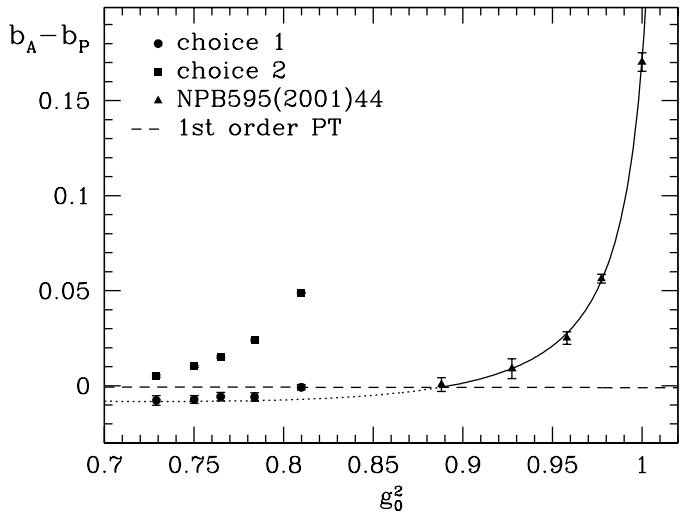,width=10.0cm}
\vspace{-2.0cm}
\caption{{\fns
Non-perturbative results for $\ba-\bP$ in the considered region of 
bare couplings for our two choices of quark masses (cf.~\eq{choices_m}), 
together with the corresponding results at larger $g_0^2$ 
from \protect\cite{impr:babp} (triangles) and the prediction from one-loop 
perturbation theory.
The solid line gives the rational fit function that was quoted 
in \protect\cite{impr:babp} to well describe the numerical simulation 
results obtained there.
}}\label{fig:bap}
\end{figure}
%
%%%%%% figure: b_m(g_0)
%
\begin{figure}[htb]
\centering
\vspace{-1.5cm}
\epsfig{file=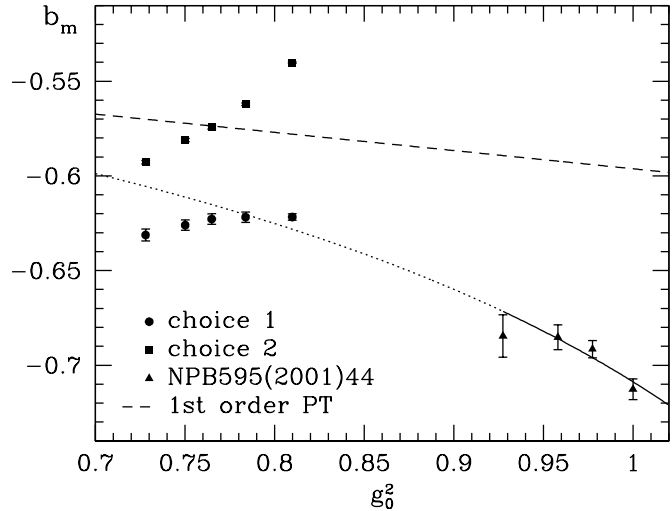,width=10.0cm}
\vspace{-2.0cm}
\caption{{\fns
The same as in \Fig{fig:bap} but for the improvement coefficient $\bm$.
In view of the leading perturbative behaviour that one expects to be 
approached in the limit $g_0^2\rightarrow 0$, the curvature seen in our 
results hints at a more complicated structure of (unknown) higher-order 
terms.  
}}\label{fig:bm}
\end{figure}
%
%%%%%% figure: Z(g_0)
%
\begin{figure}[htb]
\centering
\vspace{-1.5cm}
\epsfig{file=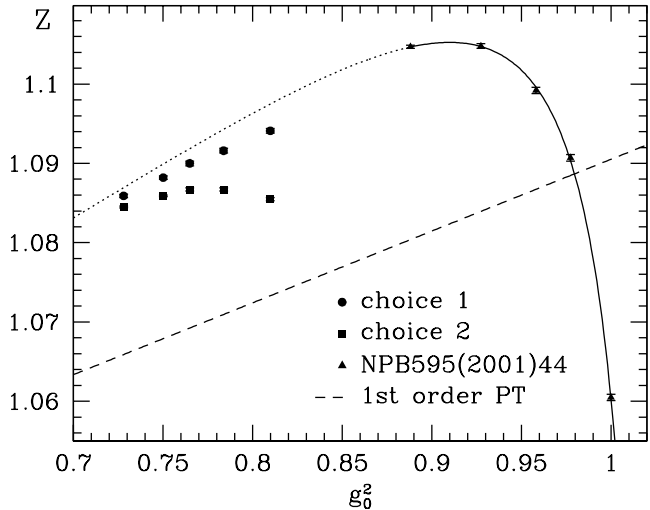,width=10.0cm}
\vspace{-2.0cm}
\caption{{\fns
The same as in \Fig{fig:bap} but for the renormalization constant $Z$.
}}\label{fig:Z}
\end{figure}
%
%%%%%% figure: Delta_Z(g_0) & Delta_b_m(g_0)
%
\begin{figure}[htb]
\centering
\vspace{-0.75cm}
\epsfig{file=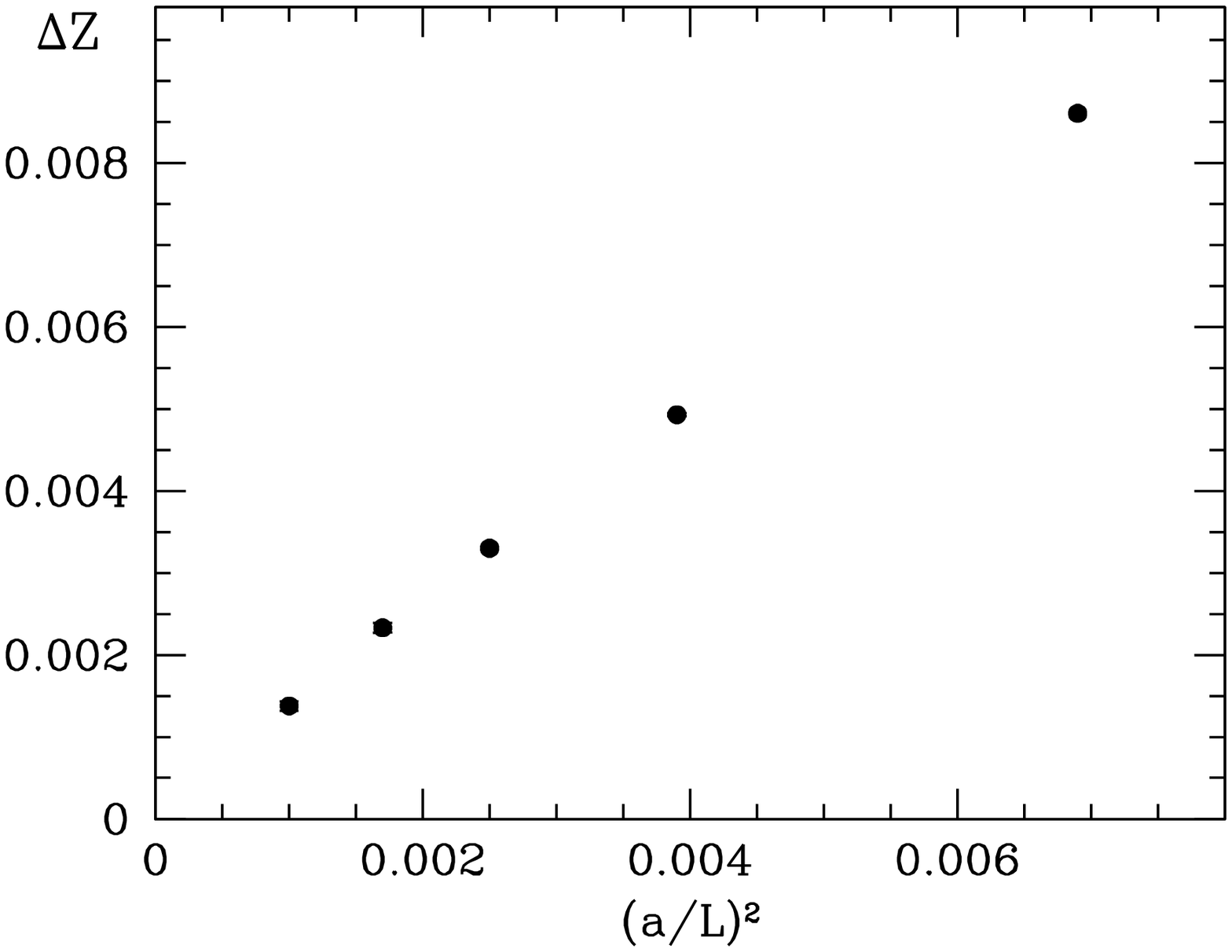,width=7.5cm}
\epsfig{file=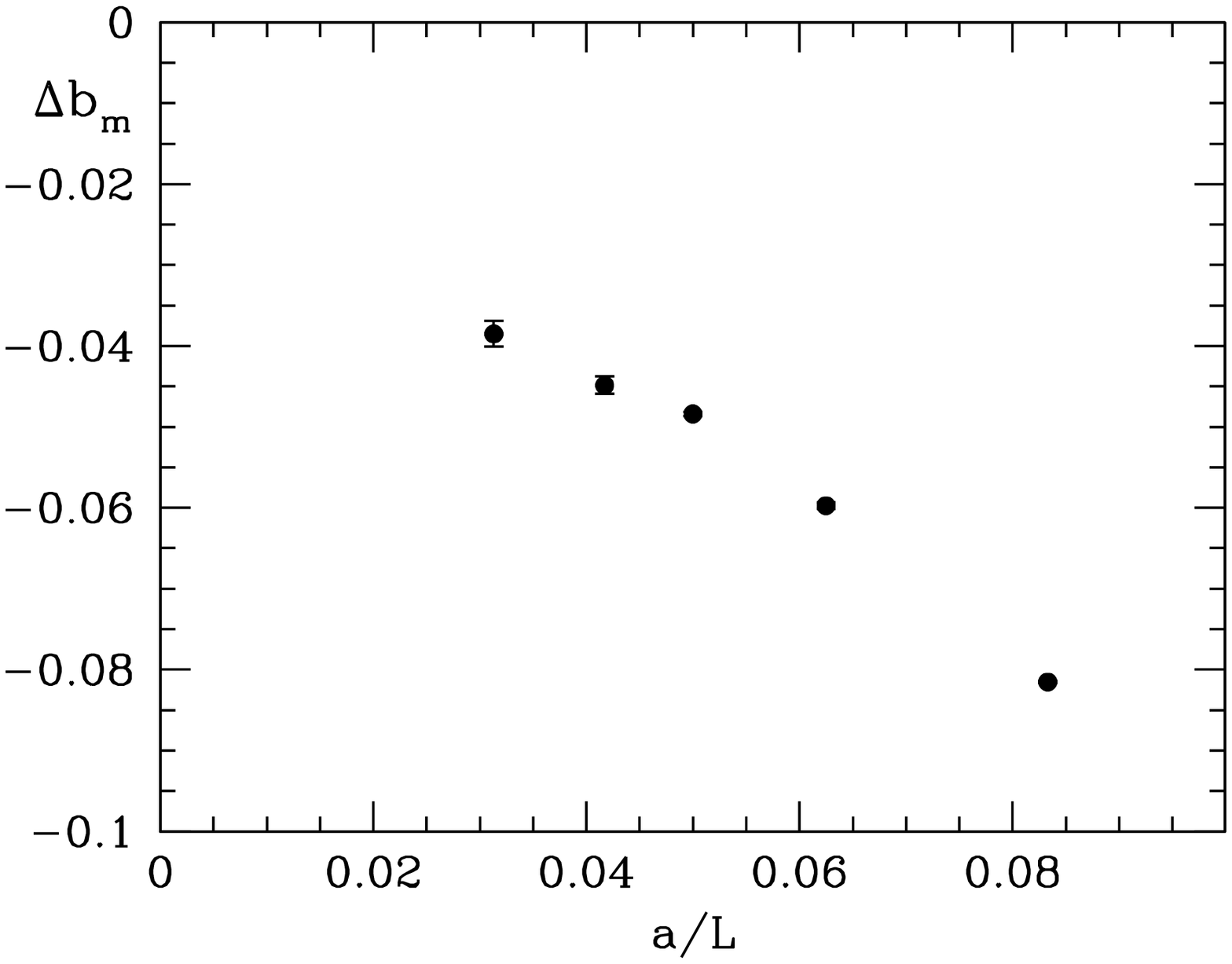,width=7.5cm}
\vspace{-1.125cm}
\caption{{\fns
Left: Difference of the results on the renormalization constant $Z$,
obtained from the two heavy quark mass choices, versus $(a/L)^2$.
Right: The same for the improvement coefficient $\bm$ where, however,
the ambiguity inherent in any improvement condition imposed is of $\Or(a)$.
}}\label{fig:delZbm}
\end{figure}
%
%%%%%% figure: L_0*Gamma_PS for z=9 and both renormalization conditions
%
\begin{figure}[htb]
\centering
\vspace{-2.25cm}
\epsfig{file=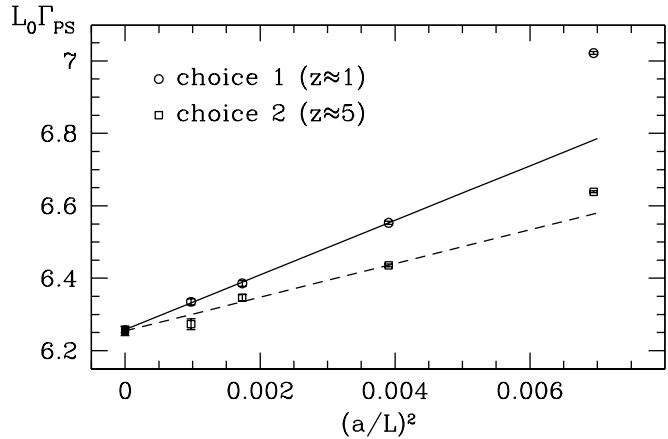,width=10.0cm}
\vspace{-2.5cm}
\caption{{\fns
$L_0\gamps(L_0,M,g_0)$, where the results on $\bm,Z$ from both 
renormalization conditions corresponding to \eq{choices_m} were used to 
keep the dimensionless RGI heavy quark mass fixed to $z=L_0M=9$ while 
varying $\beta=6/g_0^2$ and thus $a/L$.
Assuming quadratic scaling violations and discarding the coarsest lattice
in the (unconstrained) extrapolations, the continuum limits coincide.
A more thorough discussion of the continuum extrapolations of the
observables $\gamx$, ${\rm X}={\rm PS},{\rm av}$, will follow 
in \Sect{Sec_res_gamrel}.
}}\label{fig:GamPS12z9}
\vspace{-0.375cm}
\end{figure}
The Monte Carlo simulation details and the technical aspects of the 
analysis to compute the observables (\ref{estim_bap}) -- (\ref{estim_Z}) 
from the measured fermionic correlation functions are essentially the same 
as in \Ref{impr:babp}.
In particular, these secondary quantities have been averaged over the
central timeslices $L/(2a),\ldots,(T-L/2)/a$ to increase statistics, and 
their statistical errors were estimated by the jackknife method.
Another issue that deserves to be mentioned is the occurrence of the 
third mass parameter $m_{0,3}$ in the determination of $\bm$, which has to 
be set properly to ensure the required (but subtle) cancellation leading to 
the expression (\ref{estim_bm}) for the estimator $R_{\rm m}$.
As in terms of the hopping parameters this amounts to form the combination
$\kappa_3=2\kappa_1\kappa_2/(\kappa_1+\kappa_2)$, the condition
$\half(m_{0,1}+m_{0,2})-m_{0,3}=0$ can be satisfied in practice only up to 
roundoff errors, which might cause a systematic uncertainty in simulations 
with single precision arithmetics.
But in contrast to \Ref{impr:babp}, where this effect had to be taken into 
account carefully, the exact numerical value of $\kappa_3$ was now 
calculated in double precision directly in the simulation program itself so 
that a rounding error contribution (reflecting some possible remnant 
imperfection in the cancellation via $m_{0,3}$) to the error on $R_{\rm m}$ 
can be neglected here. 

Our non-perturbative results on $\ba-\bP$, $\bm$ and $Z$, which we obtained
from the numerical simulation data along these lines using improved
derivatives throughout, are also collected
in \Tabs{tab:bXzres_1} and \ref{tab:bXzres_2}.\footnote{
We note in passing that memory and processor-topology restrictions of the
APEmille parallel computers in use prevented us from directly simulating
$L/a=20$ lattices at an earlier stage of our work.
Therefore, we employed $16^3\times30$ and $24^3\times30$ lattices and a
linear interpolation in $1/L$ to arrive at the results for set C 
in \Tabs{tab:bXzres_1} and \ref{tab:bXzres_2}.
}
As a consequence of the fact that all renormalized quantities have been
fixed in units of $L$ while the ratio $L/\lmax$ and thereby also the
renormalized SF coupling are kept constant, the estimates $R_{\rm X}$ become 
smooth functions of the bare coupling, $g_0^2=6/\beta$.
This is well fulfilled in \Figs{fig:bap} -- \ref{fig:Z}, where our results 
are shown in conjunction with those at larger couplings of the earlier 
work \cite{impr:babp} together with the one-loop perturbative 
predictions \cite{impr:pap5,impr:babp}.
One observes that the estimates for $\ba-\bP$ and $Z$ corresponding to 
`choice~1' of quark masses are roughly consistent with the fit functions 
quoted in \Ref{impr:babp} for larger couplings, if one assumes them to be 
even valid far beyond the $\beta$--region where the underlying data were 
actually taken.
Nonetheless, this was not obvious from the beginning, since despite the 
quark mass values being comparable in that case the conditions defining the 
associated lines of constant physics are different.

For $\bm$ the situation is more intricate.
Here a naive prolongation of the curve from \cite{impr:babp} to weak
couplings (dottet line in \Fig{fig:bm}) does not give the right 
$g_0^2$--behaviour for this region. 
We infer this from the fact that the difference of the dotted line to our 
results does not seem to be compatible with being of $\Or(a)$.
In other words, if the improvement condition of \cite{impr:babp} were used
also in the region of weaker couplings, the points would look quite
differently, and we have to conclude that a simple approach to one-loop
perturbation theory is not an adequate representation for the continuation
of the triangles in \Fig{fig:bm}.
This underlines the importance of using particularly adapted improvement 
conditions, which may be used in the coupling range actually relevant for 
the desired application. 
The example of $\bm$ thus illustrates that our redetermination of the 
$b$--coefficients and the $Z$--factor indeed eliminates a source of 
uncontrollable error. 

On the other hand, the results from `choice~2' (i.e.~with the quark mass
$\mh$ being larger) fall significantly apart throughout, because the 
$a$--effects are generically larger in that case.
But this only reveals an expected, inevitable property of the procedure 
applied: any other estimate $R_{\rm X}$ (e.g.~stemming from a different 
choice of renormalization condition) may yield a different functional 
dependence upon $g_0^2$, but its differences are again smooth functions that 
must vanish in the continuum limit with a rate proportional to $a/L$ 
(for improvement coefficients) or even $(a/L)^2$ (for renormalization 
constants).
These intrinsic $\Or(a^n)$ ambiguities, $n=1,2$, imply that rather than a 
numerical value at some given $\beta$, the essential information lies in 
the \emph{correct $g_0^2$--dependence} of the results for the estimators 
$R_{\rm X}$, ${\rm X}={\rm AP},{\rm m},Z$, resulting from working at fixed 
physics while varying $\beta$.

To demonstrate the last statement, we also investigated a few alternative
improvement conditions, which are either realized by defining the estimators 
$R_{\rm X}$ with standard instead of improved derivatives 
(as in \cite{impr:babp}) or directly by the two different quark mass 
settings that we already have at our disposal through the choices
in \eq{choices_m}.
As an example we plot in the left part of \Fig{fig:delZbm} the difference
$\Delta Z(g_0^2)=Z(g_0^2)|_{\,\rm choice\,1}-Z(g_0^2)|_{\,\rm choice\,2}$ 
versus $(a/L)^2$, which clearly shows a linear approach towards zero.
Other cases behave similarly, e.g.~the $\Or(a)$ ambiguities for 
$\Delta\bm(g_0^2)=
\bm(g_0^2)|_{\,\rm choice\,1}-\bm(g_0^2)|_{\,\rm choice\,2}$ in the right 
part of \Fig{fig:delZbm} are found to be very small and rapidly decreasing
in magnitude as $a/L\rightarrow 0$.

As a further (and less direct) check for the universality of the continuum
limit we consider a physical quantity that depends on $\bm$ and $Z$ in a 
more implict way, namely the energy $L_0\gamps(L_0,M,g_0)$ introduced in 
\eq{gamps} of the previous section.
In fixing $z$ while computing $L_0\gamps$ for various lattice resolutions, 
the just determined results on $\bm$ and $Z$ enter via \eq{M_mqtil}, and
hence it is interesting to confront the lattice spacing dependences of
$L_0\gamps|_{\,\bm,Z:\,\rm choice\,1}$ and
$L_0\gamps|_{\,\bm,Z:\,\rm choice\,2}$ with eachother.
We did this exercise for $z=9$ (where, due to the large quark mass, 
$a$--effects are already very pronounced) and anticipate results from the 
following section in \Fig{fig:GamPS12z9} to display the two data sets and 
its continuum extrapolations linear in $(a/L)^2$.
From the nice agreement of the continuum limits\footnote{
The fact that in this case the cutoff effects in $L_0\gamps$ are larger with
$\bm,Z$ from `choice~1' (where the quark mass fixed is smaller) is not so
surprising, since via `choice~2' as improvement condition 
(where $z\approx 5$) one is closer to the line in parameter space with 
$z=9$ along which $L_0\gamps$ is computed.
}
we infer once more that our results (on $\bm$ and $Z$) 
correctly model --- within each choice of improvement condition 
separately --- the respective $g_0^2$--dependences, entailing convergence
to the continuum limit with leading corrections of $\Or(a^2)$.
\subsection{Heavy-light meson energies}
\label{Sec_res_gamrel}
Supposing the parameters $\beta$ and $\kapl$ at each $L/a$ ($=L_0/a$) to be
appropriately fixed to comply with the conditions 
$\gbsq(L_0/2)=1.8811={\rm constant}$ and $\ml=0$, \eqs{cond_gbarL0h} and 
(\ref{cond_ml}), we still have to prescribe a sequence of dimensionless 
quark mass values $z$ in order to be able to map out the heavy quark mass 
dependence of the observables $\gamx$, ${\rm X}={\rm PS},{\rm av}$, over a 
reasonable range that encloses the RGI mass scale of the b-quark itself.
To this end we split, according to the discussion around \eq{M_mqtil} 
in \Sect{Sec_strat_mqdep}, $z=L_0M$ into the product
\be
z=L_0\times h(L_0)\times\zm\times\mqh(1+\bm a\mqh)\,,\quad
\zm=\frac{Z\,\za}{\zp}\,.
\label{prod_z}
\ee
For $\bm(g_0^2)$ and $Z(g_0^2)$ we decided to use the results 
of \Tab{tab:bXzres_1} from `choice~1' of quark mass settings, mainly because
especially for $\bm$ the $g_0^2$--dependence is weaker and with the 
corresponding value of the (heavy) quark mass fixed in their determination
we more resemble the condition that in \Ref{impr:babp} was found favourable 
also from the perturbative point of view.
$\za(g_0^2)$ is known in $\Or(a)$ improved quenched QCD via the 
formula \cite{impr:pap4}
\be
\za=\frac{1-0.8496\,g_0^2+0.0610\,g_0^4}{1-0.7332\,g_0^2}\,,\quad
g_0^2\le 1\,, 
\ee
while the required values of $\zp(g_0,L_0/a)$ have already been quoted
in the last column of \Tab{tab:param} and, as was said after \eq{expr_h0}, 
the universal factor
\be
h(L_0)=1.531(14)\,,\quad L_0=\Lmax/2\,, 
\ee
could be extracted from the data published in \Ref{msbar:pap1}.
Given some value of $z$ one is aiming at, the relation between the 
subtracted bare (heavy) quark mass and the hopping parameter,
\be
a\mqh=\frac{1}{2}\left(\frac{1}{\kaph}-\frac{1}{\kapc}\right)\,,
\ee
then allows to straightforwardly solve \eq{prod_z} for $\kaph$ and, together
with $\kapl$ as quoted in \Tab{tab:param} of \Sect{Sec_strat_mqdep}, yields 
the pairs $(\kapl,\kaph)$ of hopping parameters, for which the numerical 
simulations with the computation of the heavy-light correlation functions 
have been performed.\footnote{
Due to the fact that the non-perturbative values for $\bm$ in the relevant
$\beta$--range lie around $-0.6$, the relation 
$z=L_0 h(L_0)\zm\mqh(1+\bm a\mqh)$ can not be inverted in favour of $a\mqh$ 
(and thereby $\kaph$) for arbitrarily high $z$--values.
In case of our largest $z$ ($=13.5$), for instance, this already restricts 
the possible inverse lattice spacings in \Tab{tab:gamres} to $L/a=16-32$.
}

Of course, the uncertainties to be associated with the various pieces
entering \eq{prod_z} translate into an error on $z$, which has to be taken 
into account also for any quantity regarded as a \emph{function of $z$}.
More precisely, the resulting error on $z$ consists of a $g_0$--dependent 
part and a universal, $g_0$--independent one:
while the former comes from the uncertainties of $\bm$, $Z$ and $\zp$ quoted 
in the tables plus an error (of 0.8\% at $\beta\approx 7.4$ down to 0.4\%
for $\beta\ge 7.8$ \cite{impr:pap4}) on $\za$, the latter stems from the 
overall uncertainty of 0.9\% on $h(L_0)$ in the continuum limit and hence 
has only to be added in quadrature \emph{after} a continuum extrapolation of 
the respective $z$--dependent quantity under study.\footnote{
Note that also this error can in principle be reduced further by increasing 
the precision of the continuum step scaling functions of \Ref{msbar:pap1}. 
}

The final parameter sets employed in the Monte Carlos simulations consist of 
the triples $(L/a,\beta,\kapl)$ in \Tab{tab:param} and the $\kaph$--values 
we arrived at as described before; they are listed together with the 
corresponding values of the dimensionless RGI heavy quark mass, $z$, and the 
$g_0$--dependent error part of the latter in \Tab{tab:gamres} at the
end of this section.
For the technical details of the runs to produce the numerical data on the 
SF heavy-light meson correlators, from which the logarithmic 
derivatives (\ref{gamps}) and (\ref{gamv}) are evaluated, we refer to the 
simulations reported in \Ref{zastat:pap3} (and its Appendix~A.2.~in 
particular) to non-perturbatively renormalize the static-light axial 
current.
The number of measurements in the statistical samples is comparable to what 
was accumulated to get the results in \Tabs{tab:bXzres_1} 
and \ref{tab:bXzres_2}.

\Tab{tab:gamres} also contains the numerical results for $L_0\gamps$ and 
$L_0\gamav$ at all values of $z$ that were considered.
To ease notation, we set
\be
\Omx(u,z,a/L)\equiv
L_0\gamx(L_0,M,g_0)\,\Big|_{\,\gbar^2(L_0/2)=u\,,\,L_0M=z}\,,\quad
{\rm X}={\rm PS},{\rm av}\,,
\label{def_Omega}
\ee
and denote its continuum limits as
\be
\omx(u,z)\equiv
\lim_{a/L\rightarrow 0}\Omx(u,z,a/L)\,,\quad
{\rm X}={\rm PS},{\rm av}\,.
\label{def_omega}
\ee
In \Tab{tab:gamres} both the statistical errors of $\Omx$ (in square 
brackets) as well as the combined statistical and $g_0$--dependent, 
$z$--induced uncertainties (in parentheses) are given, the latter being
obtained by including the propagation of the $g_0$--dependent part of the 
error on $z$.
As an outcome of this error analysis we find that summing up the various 
errors on the factors in \eq{prod_z} quadratically (since they are 
uncorrelated) and multiplying with the numerically estimated slopes 
$\left|\left[\,\partial\Omx/\partial z\,\right](g_0)\right|$ yields a 
contribution of about 0.3\% or less to the final uncertainties on $\Omx$.

As we work in the $\Or(a)$ improved theory, the numbers for $\Omps$ and 
$\Omav$ can now be extrapolated linearly in $a^2$ to the continuum 
limit.\footnote{
As in \cite{msbar:pap1,zastat:pap3}, the influence of the only 
perturbatively known SF-specific boundary improvement coefficients $\ct$ 
and $\cttil$ is negligible at the level of our precision.
}
However, based on the experience made in perturbation theory (see Section~5
of \Ref{zastat:pap2}) that the discretization errors primarily depend on 
the mass of the heavy quark, $aM=z\times a/L$, we expect some deviation from 
a leading linear behaviour in $a^2$ for the coarsest lattices as $z$ grows.
Since our non-perturbative data qualitatively confirm this picture, we
again adopt perturbation theory as a guide to get a rough estimate for the 
quark mass value where this deviation sets in.
In \cite{zastat:pap2} the heavy quark mass dependence of the discretization
errors in a typical matrix element of the heavy-light axial current,
similarly constructed from SF correlators as our energy observables, 
indicates a breakdown of $\Or(a)$ improvement beyond 
$(a\mhbMS\,)^2 \approx 0.2$, which with 
$\mhbMS\approx 0.7M$ \cite{msbar:pap3} approximately corresponds to 
$aM\approx 0.64$.
From this we deduce the following \emph{two-step criterium} to carefully 
perform the continuum extrapolations of $\Omx$, ${\rm X}={\rm PS},{\rm av}$, 
for the various values of $z$:
\begin{enumerate}
\item In view of the $z$--dependence of the size of the lattice artifacts,
      only allow for fits which start at and beyond the minimal $L/a$ such 
      that $aM<0.6$ approximately holds.
\item Among these, the $n$--point fit must agree with the $(n+1)$--point one 
      within errors but the former, which omits the coarsest of the lattices 
      meeting 1.~and thereby has the larger error, gives the final estimate
      of the continuum limit.
\end{enumerate}

%
%%%%%% figure: CL extrapolations of L_0*Gamma
%
\begin{figure}[htb]
\centering
\epsfig{file=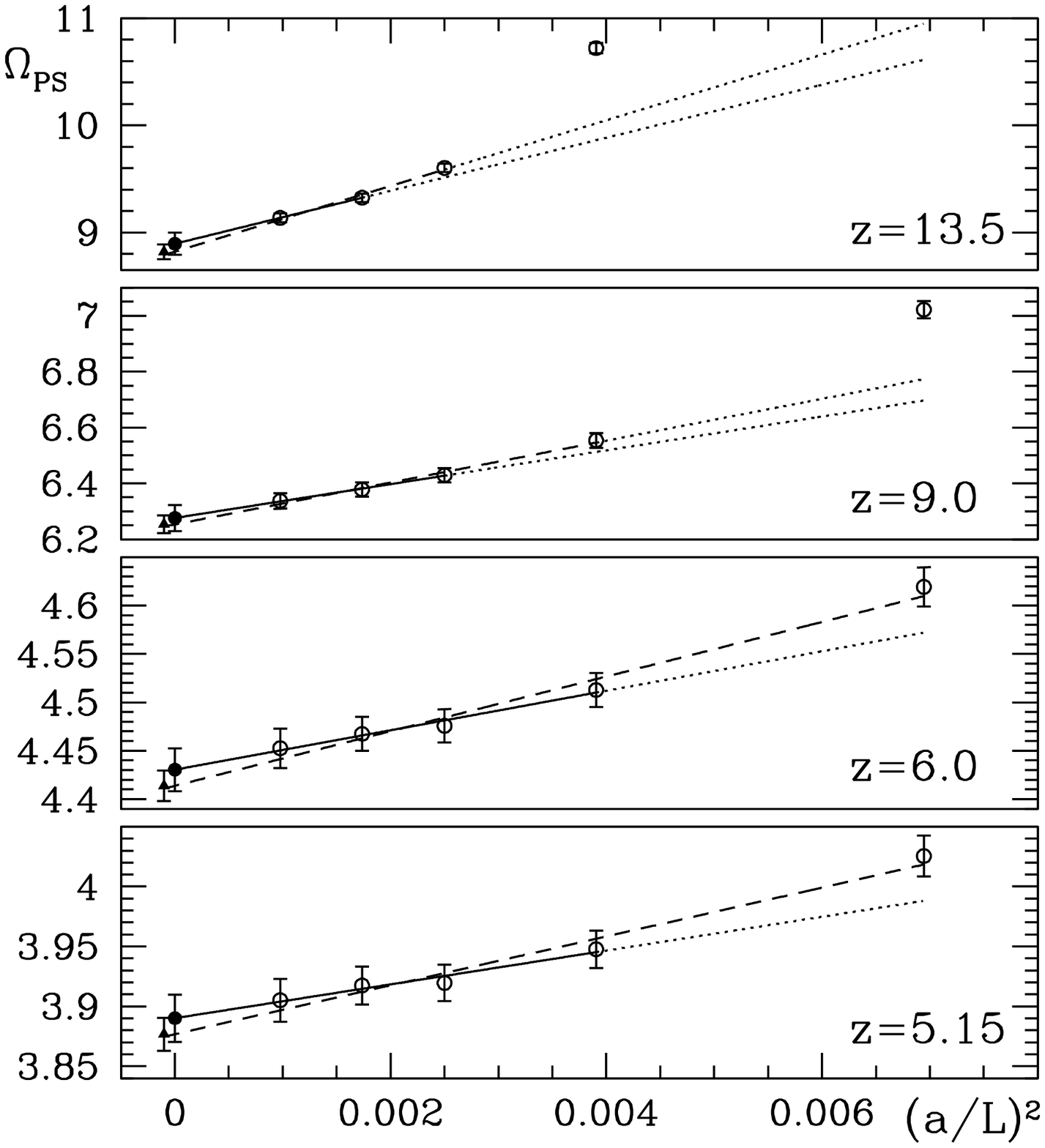,width=7.5cm}
\epsfig{file=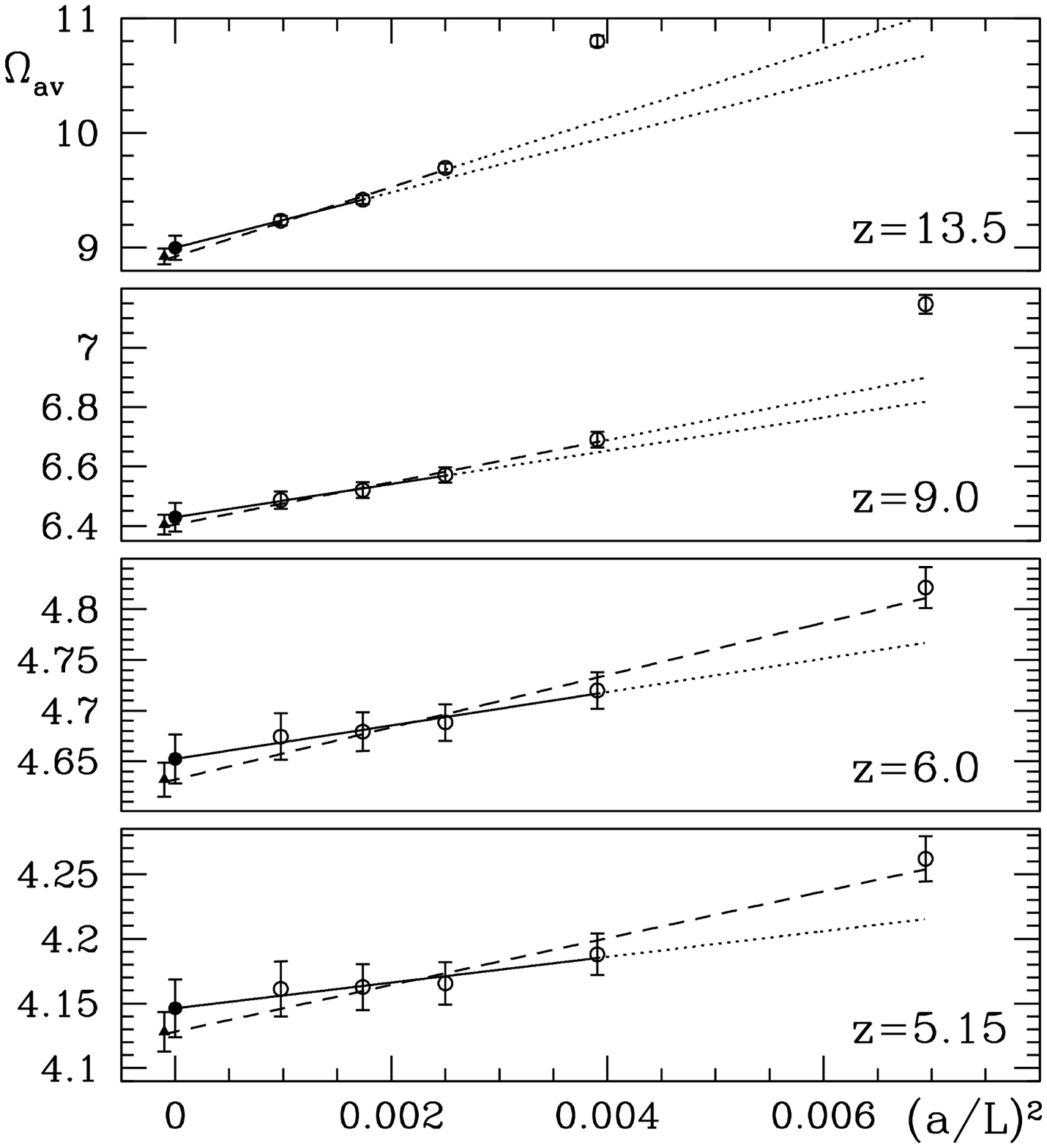,width=7.5cm}
\vspace{-0.125cm}
\caption{{\fns
Continuum extrapolations of $\Omps(1.8811,z,a/L)$ (left) and 
$\Omav(1.8811,z,a/L)$ (right) for representative values of $z$.
As explained in the text, the number of coarsest lattices to be skipped in 
the fits depends on $z$.
The final continuum limits are taken from the solid fit functions, which
are compatible with the dashed extrapolations omitting one point less.
}}\label{fig:Gamma}
\end{figure}
The resulting continuum extrapolations for the subset 
$z\in\{5.15,6.0,9.0,13.5\}$ of available $z$--values are displayed 
in \Fig{fig:Gamma}.
As it turns out when applying this fitting procedure, the continuum limits 
then have to be extracted from fits discarding the $L/a=12$ lattice in case 
of $z\in\{3.0,3.8,5.15,6.0,6.6\}$, the $L/a=12,16$ lattices in case of 
$z=9.0$ and even the $L/a=12,16,20$ lattices for the heaviest mass, 
$z=13.5$.
The numbers of \Tab{tab:gamres} in Italics are the continuum limit results 
for all $z$, based on these fits.

To further corroborate our results from this prescription, we in addition
considered an alternative fit ansatz that also accounts for a term cubic
in $a$.
Again, it is to some extent guided by the aforementioned finding 
of \Ref{zastat:pap2} that in perturbation theory the cutoff effects in the
regime of large quark masses are approximately a function of $aM$ and not of
$aM$ and $a/L$ separately.
In the $\Or(a)$ improved case at hand (and here for the example of $\Omav$,
omitting the argument $u$ for a moment), such an ansatz may therefore be 
written as: 
\be
\Omav(z,a/L)=
\omav(z)\left[\,1+\rho_2\,z^3\left(\frac{a}{L}\right)^3\,\right]
\,+\,\rho_1(z)\left(\frac{a}{L}\right)^2\,,
\ee
where by a $z$--dependent parameter $\rho_1$ we after all admit a still more 
general form for the $a^2$--term.
Performing a simultaneous fit of all available $\Omav$ data 
(no cut on $aM$), we then arrive at continuum limits in complete accordance 
with the former, having comparable or even smaller errors: 
$\omav(z)\in\{2.86(1),3.34(1),4.15(3),4.65(2),5.00(2),6.39(3),8.99(6)\}$
for $z=3.0,\ldots,13.5$.
But keeping in mind that for $aM\approx 0.6-0.8$ the perturbative 
$a$--expansion entirely breaks down \cite{zastat:pap2}, we take the 
continuum limits of \Tab{tab:gamres} from the linear extrapolations in 
$(a/L)^2$ imposing a cut on $aM$ as a safeguard against any uncontrollable 
higher-order behaviour as our final results, which within their (larger) 
errors are moreover also consistent with the values at smallest lattice 
resolution.
All told, we thus are confident that possible systematic uncertainties 
in the fitting procedure are already well covered by these estimates.

\begin{figure}[htb]
\vspace{-1.0cm}
\centering
\epsfig{file=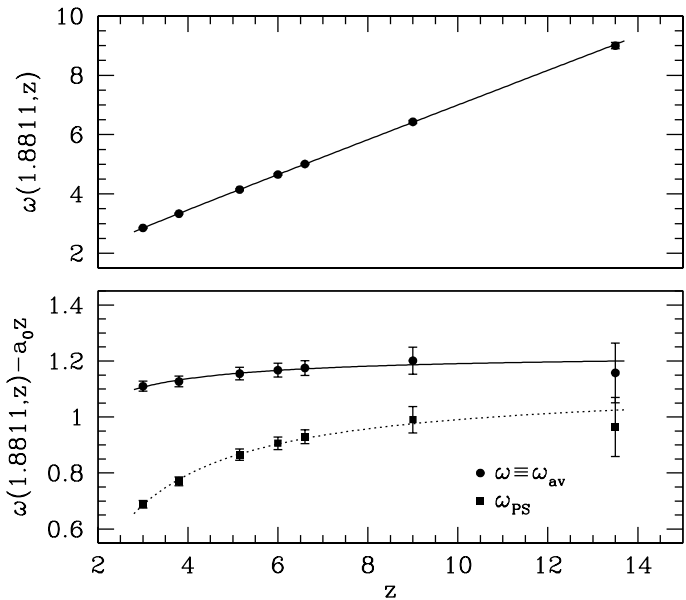,width=10.0cm}
\vspace{-1.5cm}
\caption{{\fns
Continuum limit values $\omega(1.8811,z)$ of the spin-averaged energies
$\Omega(1.8811,z,a/L)$, at fixed coupling $\gbar^2(L_0/2)=1.8811$, as a 
function of $z=L_0M$ and its fit function (top).
The bottom graph has the linear term, $a_0z$, subtracted and also includes
the pseudoscalar case, $\omps(1.8811,z)$, for comparison.
}}\label{fig:omega}
\end{figure}

For future reference (i.e.~in particular for the non-perturbative 
determination of the b-quark's mass in \Ref{mbstat:pap1}) we introduce for 
the dimensionless, spin-averaged heavy-light meson energy in small volume
the further abbreviation
\be
\Omega(u,z,a/L)\equiv\Omav(u,z,a/L)\,,\quad
\omega(u,z)\equiv\omav(u,z)\,.
\ee
To parametrize the $z$--dependence of our continuum values 
$\omega(1.8811,z)$ by a smooth fit function, we make the ansatz
$a_0 z+a_1+a_2/z$, which is justified by the theoretical expectation that
heavy-light meson correlation functions of the type studied here decay
with a leading term proportional to the heavy quark mass, up to some 
low-energy scale of $\Or(\lQCD)$ and $1/m$--corrections. 
Since the simulation data for the various $z$ at given $\beta$ were
produced on the same gauge field backgrounds, possible correlations in $z$
had to be taken into account in this fit. 
Hence we performed it on basis of the jackknife samples that were built
from the raw data and passed through the whole analysis, and we end up 
with the parametrization
\be
\omega(1.8811,z)= 
a_0 z+a_1+a_2\,\frac{1}{z}\,,\quad 
a_0=0.581 \,,\quad
a_1=1.226 \,,\quad
a_2=-0.358\,,
\label{fit_omega}
\ee
the graph of which is shown in the upper diagram of \Fig{fig:omega} to well 
represent our data.
%
%%%%%% figure: fit of omega(1.8811,z)
%
In the interval $5.2 \le z \le 6.6$, which for instance is the relevant 
$z$--range to calculate the RGI b-quark mass by means of the 
non-perturbative matching to HQET in \Ref{mbstat:pap1}, this parametrization
describes $\omega(1.8811,z)$ with a precision of about 0.5\%.
As already mentioned, a further uncertainty of 0.9\% that enters only 
indirectly through the argument $z$ of the function $\omega$ (and thus 
remains to be added in quadrature at the end) originates from the universal 
part $h(L_0)$ in \eq{prod_z} of the renormalization factor relating the 
bare to the RGI quark mass.

For later use of the result (\ref{fit_omega}) it is also necessary to have a 
numerical estimate for the derivative $\omega'(1.8811,z)$ w.r.t.~$u$, in 
order to compensate for a possible slight mismatch in the imposed condition 
of fixed renormalized coupling.
From an additional simulation with $L/a=24$ and a nearby coupling of
$\gbsq(L_0/2)=1.95$ (and assuming the $a$--effect of the derivative to be 
negligible) we found this derivative to be constant in the central region
\be
6.0\leq z\leq 6.6:\quad
\frac{\partial}{\partial u}\,\omega(u,z)\,\bigg|_{\,u=1.8811}=0.70(1)\,, 
\label{fit_omegaprime}
\ee
which is the required one in the context of the immediate application 
discussed in \cite{mbstat:pap1}.

Finally, we give as well the fit result for $\omps(1.8811,z)$ for comparison,
i.e.~if in the definition (\ref{def_Omega}) the effective energy in the 
pseudoscalar channel alone, $\gamps(L_0,M,g_0)$, would be used instead of 
the spin-averaged combination.
By an analogous analysis with the same ansatz for the fit function as 
in \eq{fit_omega} we obtain the coefficients
\be
a_0=0.587 \,,\quad
a_1=1.121 \,,\quad
a_2=-1.306 \quad (\mbox{for $\gamx=\gamps$})\,.
\ee
The fact that the absolute value of $a_2$ is evidently larger than in the 
spin-averaged case is illustrated by the shape of the subtracted fits in the 
bottom diagram of \Fig{fig:omega} and confirms our earlier claim that with 
$\gamx=\gamav$ the matching condition (\ref{cond_gam}) becomes 
\emph{independent of the coefficient of the chromomagnetic field strength 
term in the HQET Lagrangian also at first order in $1/m$.}
Therefore, the use of $\Omega\equiv\Omav\equiv L_0\gamav$ 
(and \eq{fit_omega} for its heavy quark mass dependence in the continuum 
limit) is preferable for later applications to HQET, since the $1/m$--errors 
arising at leading order will generically be smaller then.
\clearpage
%
%%%%%% table: results on L_0*Gamma and their CLs
%
\begin{table}[htb]
\centering
\begin{tabular}{lccccr@{.}lcr@{.}lr@{.}l}
\hline
  $z$ && set && $\kaph$ & \multicolumn{2}{c}{$z(g_0)$} &
& \multicolumn{2}{c}{$\Omps$} & \multicolumn{2}{c}{$\Omav$} \\
\hline
  3.0  && A && $0.129571$ & $3$&$000(12)$   && $2$&$534(11)[5]$ 
& $2$&$916(11)[6]$          \\
       && B && $0.130445$ & $3$&$000(13)$   && $2$&$485(10)[4]$ 
& $2$&$869(11)[6]$          \\
       && C && $0.130876$ & $3$&$000(13)$   && $2$&$465(10)[5]$ 
& $2$&$856(12)[8]$          \\
       && D && $0.131115$ & $3$&$000(13)$   && $2$&$471(11)[7]$ 
& $2$&$863(14)[12]$         \\
       && E && $0.131349$ & $3$&$000(13)$   && $2$&$458(13)[10]$ 
& $2$&$858(18)[16]$         \\
       && {\it CL} &&& \multicolumn{2}{c}{} && ${\it 2}$&${\it 451(13)}$ 
& ${\it 2}$&${\it 852(18)}$ \\
\hline
  3.8  && A && $0.128310$ & $3$&$800(16)$   && $3$&$094(13)[5]$  
& $3$&$408(14)[6]$          \\
       && B && $0.129552$ & $3$&$800(16)$   && $3$&$039(12)[4]$  
& $3$&$356(13)[6]$          \\
       && C && $0.130185$ & $3$&$800(16)$   && $3$&$019(12)[5]$  
& $3$&$341(13)[8]$          \\
       && D && $0.130553$ & $3$&$800(16)$   && $3$&$022(12)[7]$  
& $3$&$345(15)[11]$         \\
       && E && $0.130940$ & $3$&$800(17)$   && $3$&$011(15)[10]$ 
& $3$&$342(19)[16]$         \\
       && {\it CL} &&& \multicolumn{2}{c}{} && ${\it 3}$&${\it 002(16)}$ 
& ${\it 3}$&${\it 334(19)}$ \\
\hline
  5.15 && A && $0.126055$ & $5$&$150(21)$   && $4$&$025(17)[5]$  
& $4$&$262(17)[6]$          \\
       && B && $0.127991$ & $5$&$150(22)$   && $3$&$948(16)[5]$  
& $4$&$188(16)[6]$          \\
       && C && $0.128989$ & $5$&$150(22)$   && $3$&$920(15)[5]$  
& $4$&$166(16)[8]$          \\
       && D && $0.129586$ & $5$&$150(22)$   && $3$&$917(16)[7]$  
& $4$&$163(18)[11]$         \\
       && E && $0.130242$ & $5$&$150(23)$   && $3$&$905(18)[11]$ 
& $4$&$161(21)[16]$         \\
       && {\it CL} &&& \multicolumn{2}{c}{} && ${\it 3}$&${\it 890(20)}$ 
& ${\it 4}$&${\it 146(22)}$ \\
\hline
  6.0  && A && $0.124528$ & $6$&$000(25)$   && $4$&$619(20)[5]$  
& $4$&$821(20)[6]$          \\
       && B && $0.126967$ & $6$&$000(25)$   && $4$&$513(18)[5]$  
& $4$&$720(18)[6]$          \\
       && C && $0.128214$ & $6$&$000(25)$   && $4$&$476(17)[5]$  
& $4$&$688(18)[8]$          \\
       && D && $0.128964$ & $6$&$000(25)$   && $4$&$467(18)[8]$  
& $4$&$679(19)[11]$         \\
       && E && $0.129796$ & $6$&$000(27)$   && $4$&$453(20)[11]$ 
& $4$&$674(23)[15]$         \\
       && {\it CL} &&& \multicolumn{2}{c}{} && ${\it 4}$&${\it 430(22)}$ 
& ${\it 4}$&${\it 652(24)}$ \\
\hline
  6.6  && A && $0.123383$ & $6$&$600(27)$   && $5$&$049(22)[5]$  
& $5$&$232(22)[6]$          \\
       && B && $0.126222$ & $6$&$600(28)$   && $4$&$912(20)[5]$  
& $5$&$100(20)[6]$          \\
       && C && $0.127656$ & $6$&$600(28)$   && $4$&$866(19)[5]$  
& $5$&$059(20)[8]$          \\
       && D && $0.128518$ & $6$&$600(28)$   && $4$&$852(19)[8]$  
& $5$&$045(21)[10]$         \\
       && E && $0.129477$ & $6$&$600(30)$   && $4$&$835(22)[11]$ 
& $5$&$037(24)[15]$         \\
       && {\it CL} &&& \multicolumn{2}{c}{} && ${\it 4}$&${\it 806(24)}$ 
& ${\it 5}$&${\it 009(26)}$ \\
\hline
  9.0  && A && $0.117762$ & $9$&$000(39)$   && $7$&$022(31)[5]$  
& $7$&$146(31)[5]$          \\
       && B && $0.122987$ & $9$&$000(39)$   && $6$&$554(27)[5]$  
& $6$&$690(27)[6]$          \\
       && C && $0.125309$ & $9$&$000(38)$   && $6$&$429(25)[6]$  
& $6$&$572(26)[7]$          \\
       && D && $0.126670$ & $9$&$000(38)$   && $6$&$378(26)[8]$  
& $6$&$521(26)[10]$         \\
       && E && $0.128175$ & $9$&$000(40)$   && $6$&$337(28)[12]$ 
& $6$&$487(29)[15]$         \\
       && {\it CL} &&& \multicolumn{2}{c}{} && ${\it 6}$&${\it 277(47)}$ 
& ${\it 6}$&${\it 429(48)}$ \\
\hline
  13.5 && B && $0.113764$ & $13$&$500(69)$  && $10$&$721(47)[5]$ 
& $10$&$801(47)[5]$         \\
       && C && $0.120094$ & $13$&$500(60)$  && $9$&$603(40)[6]$ 
& $9$&$695(40)[7]$          \\
       && D && $0.122808$ & $13$&$500(58)$  && $9$&$324(38)[8]$ 
& $9$&$418(38)[9]$          \\
       && E && $0.125575$ & $13$&$500(61)$  && $9$&$136(40)[12]$ 
& $9$&$235(41)[14]$         \\
       && {\it CL} &&& \multicolumn{2}{c}{} && ${\it 8}$&${\it 89(10)}$ 
& ${\it 9}$&${\it 00(11)}$  \\
\hline \\[-3.0ex]
\end{tabular}
\caption{{\fns
Heavy quark masses $z=L_0M$ with $g_0$--dependent error and associated 
results for $\Omx(u,z,a/L)$, ${\rm X}={\rm PS},{\rm av}$, $u=1.8811$, with 
the total $g_0$--dependent part of the error in parentheses and the only 
statistical one of $\Omx$ in square brackets. 
Continuum limits (see text) are displayed in Italics.
}}\label{tab:gamres}
\end{table}
\clearpage
\section{Conclusions}
\label{Sec_concl}
In this work we have computed the non-perturbative heavy quark mass
dependence of effective energies derived from heavy-light meson correlation
functions by means of numerical simulations. 
A particular aspect of this computation is the \emph{use of a physically 
small volume} (of a linear extent of $\Or(0.2\,\fm)$): 
it is a prerequisite to treat the heavy quark flavour in lattice 
regularization as a relativistic particle and thus may be looked at as a 
`device' that serves to non-perturbatively renormalize HQET with the method 
of matching the effective theory to small-volume QCD, proposed and applied 
in \Ref{mbstat:pap1}.

Among the several improvement coefficients and renormalization constants
that are needed to accurately keep fixed the renormalized (heavy) quark mass 
while the observables of interest approach their continuum limits, we 
determined those relating the renormalized to the subtracted bare quark mass 
in a range of weaker couplings relevant here 
($7.4\lesssim\beta=6/g_0^2\lesssim 8.2$) with high precision.
Even in the region of the b-quark mass and slightly beyond we still find our 
results on the meson energies under study to be rather mildly cutoff 
dependent so that the continuum extrapolations can be well controlled.

The quenched numerical results obtained in \Sect{Sec_res_gamrel} directly 
pass into the --- for the first time entirely non-perturbative --- 
calculation of the b-quark mass in the static 
approximation of \cite{mbstat:pap1,mbstat:pap2}.
Moreover, our quantitative investigation of the quark mass dependence in 
relativistic heavy-light systems in finite volume can easily be extended to 
other heavy-light bilinears and matrix elements, which then opens the 
possibility to perform genuinely non-perturbative tests of HQET and to 
estimate the size of the $1/m$--corrections to the static theory. 
We will focus on these issues in a separate 
publication \cite{QCDvsHQET:pap2}.

Finally, considering the small-volume investigation presented here in the
light of the more general framework \cite{mbstat:pap1} of a 
non-perturbative matching between HQET and QCD, it is an obvious practical 
advantage that they can be transfered to also include dynamical quarks in 
the (hopefully near) future without requiring exceedingly large computing 
resources.
\subsection*{Acknowledgements}
This work is part of the ALPHA Collaboration research programme.
The largest part of the numerical simulations has been performed on the
APEmille computers at DESY Zeuthen, and we thank DESY for allocating 
computer time to this project as well as the staff of the computer centre 
at Zeuthen for their support.
In addition we ran on the PC cluster of the University of M\"unster a C-code 
based on the MILC Collaboration's public lattice gauge theory 
code \cite{code:MILC}, which in its version incorporating Schr\"odinger 
functional correlation functions was kindly provided to us by A.~J\"uttner.
We are also grateful to R.~Sommer for useful discussions and a critical 
reading of the manuscript.
This work was supported in part by the EU IHP Network on 
\emph{Hadron Phenomenology from Lattice QCD} under grant HPRN-CT-2000-00145.

\bibliography{lattice_ALPHA}

\begin{thebibliography}{10}

\bibitem{Back:2003ty}
BABAR, J.J. Back,
\newblock Nucl. Phys. Proc. Suppl. 121 (2003) 239, hep-ex/0308069.
%%CITATION = HEP-EX 0308069;%%

\bibitem{Yamauchi:2003rw}
Belle, M. Yamauchi,
\newblock Nucl. Phys. Proc. Suppl. 117 (2003) 83.
%%CITATION = NUPHZ,117,83;%%

\bibitem{Zoccoli:2003ih}
HERA-B, A. Zoccoli,
\newblock Nucl. Phys. A715 (2003) 280.
%%CITATION = NUPHA,A715,280;%%

\bibitem{lat02:bphys}
N. Yamada,
\newblock Nucl. Phys. Proc. Suppl. 119 (2003) 93, hep-lat/0210035.
%%CITATION = HEP-LAT 0210035;%%

\bibitem{stat:eichten}
E. Eichten,
\newblock Nucl. Phys. Proc. Suppl. 4 (1988) 170.
%%CITATION = NUPHZ,4,170;%%

\bibitem{stat:eichhill1}
E. Eichten and B. Hill,
\newblock Phys. Lett. B234 (1990) 511.
%%CITATION = PHLTA,B234,511;%%

\bibitem{Maiani:1992az}
L. Maiani, G. Martinelli and C.T. Sachrajda,
\newblock Nucl. Phys. B368 (1992) 281.
%%CITATION = NUPHA,B368,281;%%

\bibitem{lat01:mbstat}
ALPHA, J. Heitger and R. Sommer,
\newblock Nucl. Phys. Proc. Suppl. 106 (2002) 358, hep-lat/0110016.
%%CITATION = HEP-LAT 0110016;%%

\bibitem{mbstat:pap1}
ALPHA, J. Heitger and R. Sommer,
\newblock J. High Energy Phys. 0402 (2004) 022, hep-lat/0310035.
%%CITATION = HEP-LAT 0310035;%%

\bibitem{mb:roma2}
G.M. de~Divitiis et~al.,
\newblock Nucl. Phys. B675 (2003) 309, hep-lat/0305018.
%%CITATION = HEP-LAT 0305018;%%

\bibitem{fb:roma2c}
G.M. de~Divitiis et~al.,
\newblock Nucl. Phys. B672 (2003) 372, hep-lat/0307005.
%%CITATION = HEP-LAT 0307005;%%

\bibitem{QCDvsHQET:pap2}
J. Heitger et~al.,
\newblock in preparation.

\bibitem{alpha:Nf2}
ALPHA, A. Bode et~al.,
\newblock Phys. Lett. B515 (2001) 49, hep-lat/0105003.
%%CITATION = HEP-LAT 0105003;%%

\bibitem{Nf2SF:algo}
ALPHA, M. {Della Morte} et~al.,
\newblock (2003), hep-lat/0307008.
%%CITATION = HEP-LAT 0307008;%%

\bibitem{mbstat:dm_MaSa}
G. Martinelli and C.T. Sachrajda,
\newblock Nucl. Phys. B559 (1999) 429, hep-lat/9812001.
%%CITATION = HEP-LAT 9812001;%%

\bibitem{mbstat:dm_DirScor}
F.D. Renzo and L. Scorzato,
\newblock J. High Energy Phys. 0102 (2001) 020, hep-lat/0012011.
%%CITATION = HEP-LAT 0012011;%%

\bibitem{mbstat:dm_Trottier}
H.D. Trottier et~al.,
\newblock Phys. Rev. D65 (2002) 094502, hep-lat/0111028.
%%CITATION = HEP-LAT 0111028;%%

\bibitem{lat01:ryan}
S.M. Ryan,
\newblock Nucl. Phys. Proc. Suppl. 106 (2002) 86, hep-lat/0111010.
%%CITATION = HEP-LAT 0111010;%%

\bibitem{alpha:sigma}
M. {L\"uscher}, P. Weisz and U. Wolff,
\newblock Nucl. Phys. B359 (1991) 221.
%%CITATION = NUPHA,B359,221;%%

\bibitem{SF:LNWW}
M. {L\"uscher} et~al.,
\newblock Nucl. Phys. B384 (1992) 168, hep-lat/9207009.
%%CITATION = HEP-LAT 9207009;%%

\bibitem{SF:stefan1}
S. Sint,
\newblock Nucl. Phys. B421 (1994) 135, hep-lat/9312079.
%%CITATION = HEP-LAT 9312079;%%

\bibitem{impr:pap3}
ALPHA, M. {L\"uscher} et~al.,
\newblock Nucl. Phys. B491 (1997) 323, hep-lat/9609035.
%%CITATION = HEP-LAT 9609035;%%

\bibitem{lat97:marco}
ALPHA, M. Guagnelli and R. Sommer,
\newblock Nucl. Phys. Proc. Suppl. 63 (1998) 886, hep-lat/9709088.
%%CITATION = HEP-LAT 9709088;%%

\bibitem{impr:pap1}
M. {L\"uscher} et~al.,
\newblock Nucl. Phys. B478 (1996) 365, hep-lat/9605038.
%%CITATION = HEP-LAT 9605038;%%

\bibitem{impr:scalt}
ALPHA, J. Heitger,
\newblock Nucl. Phys. B557 (1999) 309, hep-lat/9903016.
%%CITATION = HEP-LAT 9903016;%%

\bibitem{alpha:SU3}
M. {L\"uscher} et~al.,
\newblock Nucl. Phys. B413 (1994) 481, hep-lat/9309005.
%%CITATION = HEP-LAT 9309005;%%

\bibitem{msbar:pap1}
ALPHA, S. Capitani et~al.,
\newblock Nucl. Phys B544 (1999) 669, hep-lat/9810063.
%%CITATION = HEP-LAT 9810063;%%

\bibitem{pot:r0}
R. Sommer,
\newblock Nucl. Phys. B411 (1994) 839, hep-lat/9310022.
%%CITATION = HEP-LAT 9310022;%%

\bibitem{pot:r0_ALPHA}
ALPHA, M. Guagnelli, R. Sommer and H. Wittig,
\newblock Nucl. Phys. B535 (1998) 389, hep-lat/9806005.
%%CITATION = HEP-LAT 9806005;%%

\bibitem{impr:babp}
ALPHA, M. Guagnelli et~al.,
\newblock Nucl. Phys. B595 (2001) 44, hep-lat/0009021.
%%CITATION = HEP-LAT 0009021;%%

\bibitem{zastat:pap3}
ALPHA, J. Heitger, M. Kurth and R. Sommer,
\newblock Nucl. Phys. B669 (2003) 173, hep-lat/0302019.
%%CITATION = HEP-LAT 0302019;%%

\bibitem{impr:ct_2loop}
ALPHA, A. Bode, P. Weisz and U. Wolff,
\newblock Nucl. Phys. B576 (2000) 517, hep-lat/9911018,
\newblock Errata: ibid. B600 (2001) 453, ibid. B608 (2001) 481.
%%CITATION = HEP-LAT 9911018;%%

\bibitem{impr:roma2_1}
G.M. de~Divitiis and R. Petronzio,
\newblock Phys. Lett. B419 (1998) 311, hep-lat/9710071.
%%CITATION = HEP-LAT 9710071;%%

\bibitem{impr:losalamos}
T. Bhattacharya et~al.,
\newblock Phys. Rev. D63 (2001) 074505, hep-lat/0009038.
%%CITATION = HEP-LAT 0009038;%%

\bibitem{pot:r0_silvia}
S. Necco and R. Sommer,
\newblock Nucl. Phys. B622 (2002) 328, hep-lat/0108008.
%%CITATION = HEP-LAT 0108008;%%

\bibitem{pot:r0_largebeta}
M. Guagnelli, R. Petronzio and N. Tantalo,
\newblock Phys. Lett. B548 (2002) 58, hep-lat/0209112.
%%CITATION = HEP-LAT 0209112;%%

\bibitem{impr:pap5}
ALPHA, S. Sint and P. Weisz,
\newblock Nucl. Phys. B502 (1997) 251, hep-lat/9704001.
%%CITATION = HEP-LAT 9704001;%%

\bibitem{impr:pap4}
ALPHA, M. {L\"uscher} et~al.,
\newblock Nucl. Phys. B491 (1997) 344, hep-lat/9611015.
%%CITATION = HEP-LAT 9611015;%%

\bibitem{zastat:pap2}
ALPHA, M. Kurth and R. Sommer,
\newblock Nucl. Phys. B623 (2002) 271, hep-lat/0108018.
%%CITATION = HEP-LAT 0108018;%%

\bibitem{msbar:pap3}
ALPHA \& UKQCD, J. Garden et~al.,
\newblock Nucl. Phys. B571 (2000) 237, hep-lat/9906013.
%%CITATION = HEP-LAT 9906013;%%

\bibitem{mbstat:pap2}
ALPHA,
\newblock in preparation.

\bibitem{code:MILC}
http://www.physics.indiana.edu/\~{}sg/milc.html.

\end{thebibliography}
\bibliographystyle{h-elsevier3}
\end{document}